\documentclass[final,5p,times,twocolumn,compress]{elsarticle}

\usepackage{graphicx}% Include figure files
\usepackage[utf8]{inputenc}
\usepackage{subfigure}
\usepackage[T1]{fontenc}
\UseRawInputEncoding
\usepackage{bibentry}
\usepackage{dcolumn}% Align table columns on decimal point
\usepackage{bm}% bold math
\usepackage[mathlines]{lineno}% Enable numbering of text and display math
\graphicspath{{figure/}}
\usepackage{array}
\usepackage{braket}
\usepackage{diagbox}
\usepackage{amsmath,amssymb,amsfonts}%
\usepackage{amsmath}
\usepackage{appendix}
\usepackage{multirow}
\usepackage{bm,color,bbm}

\newcommand{\blk}{\color{black}}

\usepackage{float}
\usepackage{dblfloatfix} 
\newcolumntype{.}{D{.}{.}{-1}}
\usepackage{amsopn}
\usepackage[colorlinks,linkcolor=blue, anchorcolor=blue, urlcolor=blue, citecolor=blue]{hyperref}
\usepackage{epstopdf}
\usepackage{natbib}
\usepackage{siunitx}
\setcitestyle{super,comma,open={},close={}}

\usepackage{hyperref}	
\hypersetup{
	colorlinks = true,
	citecolor = {blue},
	linkcolor = {red},
	pdftitle = {Security Risks of VOA-Induced Luminescence in Chip-Based quantum key distribution}, 
	pdfauthor = {Zijian Li et al.} 
}
\usepackage{lineno}

\usepackage[utf8]{inputenc} 
\usepackage[T1]{fontenc}   
\makeatletter
\def\ps@pprintTitle{%
	\let\@oddhead\@empty
	\let\@evenhead\@empty
	\let\@oddfoot\@empty
	\let\@evenfoot\@empty
}
\makeatother

\begin{document}

\begin{frontmatter}

\title{Security Risks of VOA-Induced Luminescence in Chip-Based quantum key distribution}

\author[label1]{Zijian Li}
\author[label1]{Chenyu Xu}
\author[label2,labe3]{Xin Hua}
\author[label1]{Yongqiang Du}
\author[label1]{Xin Liu}
\author[label1]{Tao Lin}
\author[label2,labe3]{Xi Xiao}
\author[label1]{Kejin Wei}
%\ead{kjwei@gxu.edu.cn}

\address[label1]{Guangxi Key Laboratory for Relativistic Astrophysics, School of Physical Science and Technology, Guangxi University, Nanning 530004, China.}

\address[label2]{National Information Optoelectronics Innovation Center (NOEIC), Wuhan 430074,  China.}       

\address[labe3]{State Key Laboratory of Optical Communication Technologies and Networks, China Information and Communication Technologies Group Corporation (CICT), 430074 Wuhan, China}        

\begin{abstract} 
	
Integrated photonics is widely regarded as a key enabler for scalable quantum key distribution (QKD), offering compactness, stability, and compatibility with semiconductor fabrication. Despite rapid advances in chip-based QKD, the implementation security of integrated photonic components remains insufficiently understood.
Here we present the first systematic study of an implementation-level security vulnerability associated with \textit{p--n} junction--based variable optical attenuators (VOAs), a ubiquitous component in integrated QKD transmitters. We theoretically and experimentally demonstrate that electrically biased \textit{p--n} junction VOAs emit spontaneous luminescence. Using a single-photon--sensitive spectral measurement technique, we identify the emission wavelength to be centered around 1107~nm, well separated from the C-band quantum signals.
This spectral separation gives rise to a previously unrecognized wavelength-resolved side channel, enabling potential wavelength-splitting attacks without directly disturbing the encoded quantum states. By incorporating the measured luminescence into a quantitative security analysis, we show that even extremely weak emission can lead to non-negligible information leakage.
Our findings reveal a fundamental and previously overlooked security risk in photonic integrated QKD systems and highlight the necessity of security-aware device design for future integrated quantum communication technologies.

\end{abstract}

\end{frontmatter}

\section{Introduction}

Quantum key distribution (QKD) exploits fundamental principles of quantum mechanics to enable information-theoretically secure key exchange between remote parties~\cite{bennett1984,gisin2002}. Over the past decades, remarkable progress has been achieved, including long-distance fiber transmission exceeding 1,000~km~\cite{liuYang2023experimental1000,liuYang2023finite1000}, gigahertz-scale clock rates~\cite{YuanZhiliang2018,boaron2018,weiKejin2020,grunenfelder2020,liwei2023,grunenfelder2023,sax2023,chenZhaoYuan2025,linZhihao2025,zhangGuoWei2025}, and the deployment of metropolitan-scale quantum networks~\cite{wangShuang2010,sasaki2011,liaoShengKai2018,dynes2019,chenYuAo2021,chenTengYun2021,krvzivc2023,huangChunfeng2024,yanWenhan2025,pittaluga2025,liuJingyang2025,guanRui2025}. These advances demonstrate the growing technological maturity of QKD and its potential for real-world deployment. However, large-scale adoption remains constrained by practical considerations such as system stability, footprint, and cost~\cite{liuQiang2022,luoWei2023,labonte2024,Zhangheqian2025}.

Integrated photonics has emerged as a leading platform to address these challenges by enabling compact, stable, and scalable QKD implementations compatible with mature semiconductor fabrication technologies~\cite{silverstone2016,wangJianwei2020}. A wide range of QKD protocols have now been successfully realized on photonic chips, including decoy-state BB84~\cite{maChaoxuan2016,sibson2017,bunandar2018,paraiso2019,avesani2021,zhangGaolu2021,beutel2022,liXiao2022,duYongqiang2023,weiKejin2023}, continuous-variable QKD~\cite{zhangGong2019,bianYiming2024,Aldama2025}, and measurement-device-independent QKD~\cite{caoLin2020,liWei2021,zhengXiaodong2021,duHan2024}. These demonstrations establish integrated photonics as a promising route toward scalable quantum communication infrastructures.

Despite this rapid progress, the security of chip-based QKD systems is still predominantly assessed at the protocol level, often assuming idealized device behavior. In practice, integrated photonic platforms combine quantum and classical functionalities within tightly confined structures, where non-ideal physical processes may give rise to unintended information leakage. Recent studies have revealed several implementation-level vulnerabilities in integrated QKD systems, including state-preparation loss~\cite{liChenyang2018,Yepeng2022}, optical back-reflections~\cite{tanHao2021,johlinger2024}, light-injection--induced effects in modulators~\cite{yePeng2023,hanLiying2023,tengJun2023,luFengYu2023,wangYiliang2025}, side channels associated with electronic power consumption~\cite{zhengYi2021}, and carrier-induced fluctuations in integrated phase modulators~\cite{liLang2021}. These findings indicate that device-level physical mechanisms, if overlooked, can fundamentally undermine the security assumptions of QKD protocols.

Here, we identify and investigate a previously unreported implementation security vulnerability associated with \textit{p--n} junction--based variable optical attenuators (VOAs), which are widely used for intensity control in integrated QKD transmitters. We theoretically predict and experimentally demonstrate that electrically biased \textit{p--n} junction VOAs emit unintended luminescence originating from the junction structure. By developing a single-photon--level spectral characterization technique, we determine that the emission is centered near 1107~nm, well separated from the C-band wavelengths used for quantum signal transmission. This spectral separation opens a wavelength-resolved side channel that can be exploited by an eavesdropper through wavelength-splitting attacks. Our security analysis further shows that even extremely weak luminescence can result in non-negligible information leakage. These results reveal a fundamental and previously overlooked security risk in integrated QKD systems and highlight the necessity of security-aware design principles for future photonic quantum communication technologies.

The remainder of this paper is organized as follows. Section~\ref{section2} presents the theoretical analysis of the \textit{p--n} junction VOA and the physical origin of its luminescence. Section~\ref{section3} describes the experimental characterization at the single-photon level. Section~\ref{section4} analyzes the resulting security threats and quantifies the information leakage under different deployment scenarios. Section~\ref{section7} concludes the paper.
\section{Physical model of silicon-based VOAs in a chip-based QKD system}\label{section2}
In this section, we describe the physical model of silicon-based VOAs in a chip-based QKD system. Then, we show that the physical mechanism of VOAs exhibits not only an attenuation effect but also parasitic luminescence, which would pose a potential security loophole to QKD systems.
\vspace{-0.2cm}
\subsection{Physical structure and operating principle}\label{VOA_Attenuation}
\begin{figure}[h]
	\centering
	\includegraphics[width=1\linewidth]{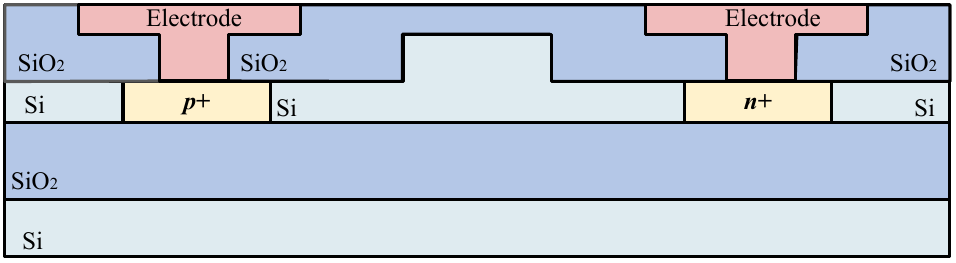}
	\caption{Cross-sectional schematic of a silicon photonic VOA on an SOI platform (not to scale), showing the Si rib waveguide, \textit{p}+/\textit{n}+ regions, $\mathrm{SiO_2}$ cladding, metal electrodes, and the Si substrate.
	}\label{VOA_Structural}
\end{figure}

Fig.~\ref{VOA_Structural} illustrates the cross-section of a typical VOA, as commonly realized in platforms from foundries such as AIM~\cite{AIMPhotonics_PDK}, CompoundTek~\cite{CompoundTek_PDK} and AMF~\cite{AdvMfg_PDK}, fabricated using a standard silicon photonics process. \blk It can be seen that a silicon-based VOA consists of a silicon substrate, a silicon dioxide buried layer, and a central silicon rib waveguide with a top width of several hundred nanometers.  Heavily doped $p^{+}$ and $n^{+}$ regions are formed in the slab regions on both sides of the rib waveguide via selective ion implantation. These doped regions are separated by several micrometers to form an intrinsic silicon waveguide region, thereby avoiding high absorption losses caused by dopants incorporated directly into the central waveguide region. To establish external electrical contact, metal electrodes are connected to these doped regions, collectively forming a lateral $p$-$i$-$n$ diode.

When a forward bias voltage is applied to the \textit{p-i-n} diode, carriers are injected into the intrinsic waveguide region. This injection of carriers modifies the refractive index and the absorption coefficient of the silicon. These changes are described by the plasma dispersion effect~\cite{Soref1987}:
\begin{equation}
	\Delta n = -\frac{q^{2}\lambda^{2}}{8\pi^{2}c^{2}\varepsilon_{0}n_{0}}\left(\frac{\Delta N_{e}}{m_{ce}^{*}}+\frac{\Delta N_{h}}{m_{ch}^{*}}\right), \label{refractive_index}
\end{equation}
\begin{equation}
	\Delta\alpha=\frac{q^{3}\lambda^{2}}{4\pi^{2}c^{3}\varepsilon_{0}n_{0}}\left[\frac{\Delta N_{e}}{\left(m_{ce}^{*}\right)^{2}\mu_{n}}+\frac{\Delta N_{h}}{\left(m_{ch}^{*}\right)^{2}\mu_{p}}\right].\label{absorption_coefficient}
\end{equation}
Here \(q\) is the electron charge; \(\lambda\) is the wavelength; \(\varepsilon_{0}\) is the permittivity of free space; \(n_{0}\) is the refractive index of intrinsic silicon; \(\Delta N_{e}\) and $\Delta N_{\mathrm{h}}$ are the concentration of electrons and holes respectively; $m_{\mathrm{ce}}^{*}$ and $m_{\mathrm{ch}}^{*}$ are the effective mass of electrons and holes respectively; $\mu_{\mathrm{n}}$ and $\mu_{\mathrm{p}}$ are the carrier mobility of electrons and holes respectively; $\Delta n$ and $\Delta \alpha$ respectively represent the change of refractive index and absorption coefficient.

In the C-band commonly used for QKD, for example at a wavelength of 1550~nm, the changes in refractive index and absorption coefficient induced by the plasma dispersion effect can be calculated using the well-known Soref empirical formulas:
\begin{equation}
	\Delta n = - \left[8.8\times 10^{-22} \Delta N_{\mathrm{e}} + 8.5\times 10^{-18} (\Delta N_{\mathrm{h}})^{0.8}\right],
\end{equation}
\begin{equation}
	\Delta\alpha = 8.5\times 10^{-18}\Delta N_{e}+6.0\times 10^{-18}\Delta N_{h}.\label{Soref_alpha}
\end{equation}
The induced change in the absorption coefficient, $\Delta\alpha$, results in optical attenuation as the signal propagates through the waveguide. For the intrinsic region of length $L$, the total attenuation ($Att.$) in decibels (dB) is given by:
\begin{equation}
	\begin{aligned}
		Att.=-10\log&\left(\frac{P}{P_0}\right)=-10\log\left(\frac{P_0e^{-\Delta\alpha L}}{P_0}\right)\\&=10\Delta\alpha\cdot\log(e)\cdot L ,
	\end{aligned}
\end{equation}
where $P$ and $P_0$ are the output and input optical power, respectively. 

\subsection{Modeling VOA's electroluminescence}\label{Anomalous_Luminescence}		
The VOA's attenuation function, as previously described, relies on a forward-biased \textit{p-i-n} structure. This basic structure is shared by many silicon photonic and optoelectronic devices, including phase modulators, photodiodes, and light-emitting diodes (LEDs). While their design parameters determine the primary functions, these devices are essentially \textit{p-n} junctions operated under an external voltage bias. Consequently, the shared basic structure means that the operation intended to achieve the primary function also induces other physical processes, manifesting as parasitic effects. For instance, as shown in Eqs.~(\ref{refractive_index}) and (\ref{absorption_coefficient}), carrier injection---primarily intended to modulate the absorption coefficient in a VOA---unavoidably alters the refractive index, introducing an unintended phase shift. Conversely, in phase modulators, the modulation of the refractive index via carrier injection is intrinsically accompanied by a change in the absorption coefficient, resulting in optical loss~\cite{liChenyang2018}.

Meanwhile, given this shared physical architecture, the internal physical processes in a forward-biased VOA fundamentally resemble those of an LED. Consequently, the carrier injection employed for optical attenuation in the VOA inevitably leads to electron--hole recombination. Since this recombination is the mechanism responsible for light emission in LEDs, we theoretically predict that silicon VOAs should also exhibit electroluminescence (EL) as a parasitic effect. To the best of our knowledge, however, EL from silicon VOAs operated under typical conditions has not yet been reported. This is likely due to the indirect bandgap of silicon, for which the recombination probability is considered low. As a result, the emitted light is extremely weak---below the sensitivity of standard photodiodes---and requires single-photon detectors for observation.

\begin{figure}[h!]
	\centering
	\includegraphics[width=1\linewidth]{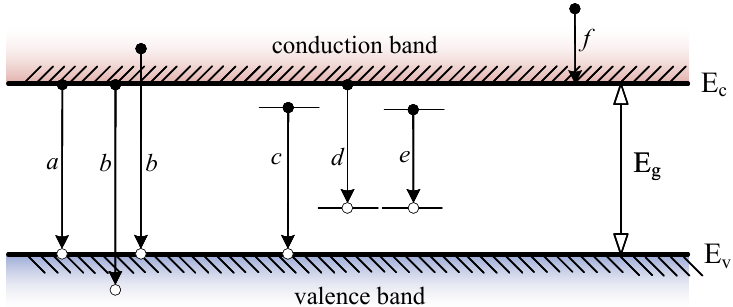}
	\caption{Schematic diagram of the main electron transition processes in a semiconductor. $\mathrm{E_c}$, $\mathrm{E_v}$, and $\mathrm{E_g}$ denote the conduction band, the valence band, and the band gap energy, respectively. Processes~$a$ and~$b$ represent intrinsic interband transitions, while $c$, $d$, and~$e$ correspond to transitions mediated by impurities or defects. Process~$f$ depicts an intraband transition.}\label{transition_ways}
\end{figure}

To elucidate the physical origin of this EL, we consider the microscopic radiative processes in silicon. Fundamentally, light emission in semiconductors such as silicon arises from the radiative recombination of injected carriers. In this process, an electron that has been excited to a higher energy level recombines radiatively with a hole by transitioning to a lower energy level, releasing the excess energy as a photon. Specifically, this radiative recombination can occur through several pathways, as illustrated in Fig.~\ref{transition_ways}: (i) interband transitions between the valence and conduction bands (e.g., processes~$a$ and~$b$, known as intrinsic luminescence); (ii) transitions involving impurities or defects (e.g., processes~$c$, $d$, and~$e$); and (iii) intraband transitions (e.g., process~$f$). Semiconductor LEDs exploit these mechanisms, where light emission is driven by forward-bias carrier injection---the same fundamental process that induces optical absorption and thereby attenuation in VOAs. This shared structural and mechanistic principle strongly suggests that light emission is an unavoidable byproduct of attenuation in VOAs.

However, as an indirect bandgap semiconductor, silicon has an extremely low intrinsic radiative efficiency (typically $10^{-6}$--$10^{-5}$ at room temperature~\cite{helm2005,canham2020}). In high-quality commercial silicon waveguides, where crystal defects and impurities are minimized, the efficiencies of impurity-assisted or intraband pathways are often even lower. While the resulting EL is negligible for classical communications, in QKD systems operating at the single-photon level even such weak emission can become a non-negligible source of background noise. To evaluate the impact of this noise, it is essential to identify the dominant radiative process. Since different mechanisms yield distinct emission spectra, the central wavelength of the EL serves as a key signature for determining the dominant luminescence mechanism.

The central wavelength $\lambda_c$ of the intrinsic luminescence (pathway~(i)) is determined by the material's bandgap energy $E_g$ according to
\begin{equation}
	\lambda_c = \frac{hc}{E_g},
\end{equation}
where $h$ is Planck's constant and $c$ is the speed of light in vacuum. For silicon, with a bandgap of $E_g \approx 1.12~\text{eV}$~\cite{seiferth1994,sze2021} at room temperature, this corresponds to an intrinsic emission wavelength of approximately $1107~\text{nm}$.

For transitions involving impurities and defects (pathway~(ii)), the emission wavelength depends on the energy levels introduced by the impurities and defects, which are typically located within the bandgap. Notably, in silicon VOAs with \textit{p--n} structures, such impurities and defects mainly originate from the high concentration of dopants in the $p^{+}$ and $n^{+}$ regions (e.g., boron or phosphorus). Consequently, luminescence with peaks typically centered around 1300--1500~nm is observed~\cite{sveinbjornsson1997,liSi2015,sobolev2016}. 

Finally, intraband transitions (pathway~(iii)) involve lower-energy processes within the same band, resulting in even longer wavelengths. In silicon, these processes typically produce broad infrared emission beyond $2000~\text{nm}$~\cite{casalino2010}.

\section{Experimental Verification}\label{section3}
\vspace{0.25cm}
We experimentally investigate the existence and physical origin of parasitic EL in silicon \textit{p--n} junction VOAs. The experiments are designed to address three key questions: (i) whether electrically biased VOAs emit detectable optical radiation, (ii) how the emission depends on the electrical operating conditions, and (iii) which radiative mechanism dominates the observed emission. Together, these measurements provide direct and unambiguous evidence for unintended EL in integrated VOAs.
\begin{figure*}[ht!]
	\centering
	\includegraphics[width=1\linewidth]{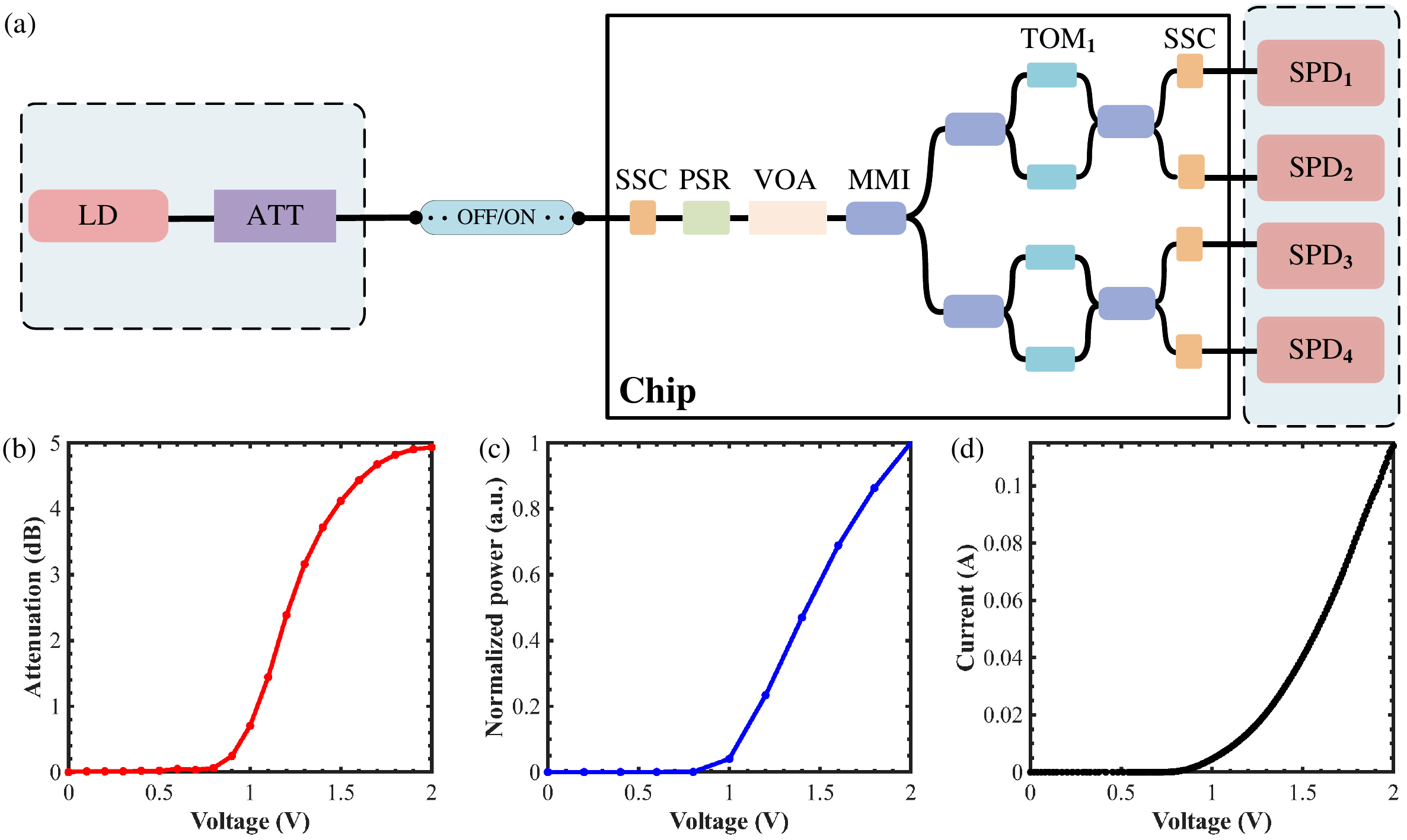}
	\caption{Schematic diagram of the experimental setup. The setup comprises off-chip optical components and an integrated test chip. The off-chip section (shown within the dashed box) includes a laser diode (LD), an off-chip attenuator (ATT), and four single-photon detectors (SPDs). The integrated test chip section (whose layout is shown in the solid box) integrates the spot-size converter (SSC), polarization splitter-rotator (PSR), thermo-optic modulator (TOM), multimode interferometer (MMI), and variable optical attenuator (VOA). The optical path from the laser to the test chip can be switched off or on (labeled as OFF/ON) depending on the measurement requirements. (b) Measured optical attenuation, (c) normalized emitted power, and (d) current of the VOA as a function of the applied voltage. The filled circles represent the measured data, and the solid lines are guides to the eye connecting the data points.}\label{VOA_Experimental_setup}	
\end{figure*}
Figure~\ref{VOA_Experimental_setup}(a) shows the experimental configuration and the silicon photonic chip used in this study. The device under test is a carrier-injection VOA based on a forward-biased $p$-$i$-$n$ junction, a widely adopted architecture for on-chip optical power control. The VOA has an interaction length of approximately \SI{200}{\micro\meter} and is monolithically integrated with passive photonic components to enable spectral characterization of the emitted light.

The chip was fabricated on a standard \SI{200}{\milli\meter} silicon-on-insulator (SOI) wafer with a \SI{220}{\nano\meter}-thick silicon device layer and a \SI{3}{\micro\meter} buried oxide layer, using a commercial \SI{90}{\nano\meter} CMOS-compatible silicon photonics process (CompoundTek platform). All components were implemented using a standard process design kit (PDK), ensuring that they are representative of typical foundry implementations for practical integrated QKD transmitters.

Light is coupled into the chip through a spot-size converter (SSC) and routed through a polarization-splitting rotator (PSR) to ensure transverse-electric (TE) mode propagation. The optical signal then passes through the VOA under test. Downstream of the VOA, a multimode interference (MMI) coupler splits the light into two arms that feed Mach--Zehnder interferometers (MZIs). These interferometers serve as wavelength-sensitive analyzers, enabling indirect measurement of the central wavelength of the emitted EL (see ~\ref{Measurement Method_Wavelength_Center} for further details). The output ports are coupled off-chip via SSCs and detected by four InGaAs single-photon detectors (WT-SPD2000, Qasky Co., Ltd.).

A pulsed laser operating at  \SI{1550.82}{\nano\meter} with a repetition rate of \SI{50}{\mega\hertz} and a pulse width of \SI{20}{\nano\second} was used as the input source. An off-chip variable attenuator reduced the optical power to the single-photon level prior to chip coupling. To isolate parasitic emission from transmitted laser light, the optical input could be independently switched on or off during the measurements. In particular, all EL measurements were performed with the input laser blocked, such that any detected photons originate exclusively from the electrically biased VOA.
\vspace{0.2cm}
\subsection{Characterization of the on-chip VOA} \label{Characterization_VOA}	
\vspace{0.1cm}	
We first characterize the electrical and optical behavior of the on-chip VOA to establish its operating regime and to provide a reference for subsequent luminescence measurements. All measurements were performed using the experimental setup shown in Fig.~\ref{VOA_Experimental_setup}(a).

To evaluate the attenuation performance, light from a laser diode was attenuated off-chip to the single-photon level and coupled into the device. A forward bias voltage applied to the VOA was swept from 0 to \SI{2}{\volt} in steps of \SI{0.1}{\volt}. For each voltage setting, photon count rates at all output ports were recorded using single-photon detectors and summed to obtain the total transmitted count rate. The optical attenuation at voltage $U$ is defined as
\begin{equation}
	\mathrm{Att.}(U) = -10 \log_{10} \left[ \frac{C(U)}{C(0)} \right],
\end{equation}
where $C(U)$ and $C(0)$ denote the total photon count rates measured at bias voltages $U$ and $0$, respectively.

The measured attenuation curve is shown in Fig.~\ref{VOA_Experimental_setup}(b). The attenuation remains below \SI{0.1}{\decibel} for voltages up to approximately \SI{0.8}{\volt}, followed by a rapid increase, reaching about \SI{4.9}{\decibel} at \SI{2}{\volt}. This behavior is consistent with previously reported carrier-injection VOAs~\cite{nishi2010,yuanPei2018,huangYuming2023} and confirms that the device operates in a standard forward-bias attenuation regime.

To directly probe unintended optical emission, the input laser was completely blocked and the VOA was forward biased from 0 to \SI{2}{\volt} in steps of \SI{0.2}{\volt}. Photon count rates were measured at each output port and converted to optical power by correcting for the detector efficiencies. The total emitted power, obtained by summing all output contributions, is plotted in Fig.~\ref{VOA_Experimental_setup}(c). The emission remains negligible at low voltages but increases sharply beyond approximately \SI{0.8}{\volt}, providing direct experimental evidence of parasitic EL during normal VOA operation.

We further characterized the electrical properties of the VOA by measuring its current-voltage ($I$-$V$) response. The forward bias voltage was swept from 0 to \SI{2}{\volt} in steps of \SI{0.01}{\volt}, and the corresponding current was recorded. The resulting $I$-$V$ curve, shown in Fig.~\ref{VOA_Experimental_setup}(d), exhibits the characteristic exponential behavior of a forward-biased \textit{p--n} junction. A detailed confirmation of the junction structure is provided in~\ref{Confirmation_PN}.

A comparison of Figs.~\ref{VOA_Experimental_setup}(b)-(d) reveals a clear correlation between injected current, optical attenuation, and emitted optical power. All three quantities increase slowly below \SI{0.8}{\volt} and rise rapidly once this threshold is exceeded. This behavior is naturally explained by the carrier-injection dynamics of the \textit{p--i--n} junction: at low bias, the voltage drop across the intrinsic region limits carrier density and current, whereas at higher bias, efficient carrier injection leads to a rapid increase in free-carrier concentration. As discussed in Sec.~\ref{section2}, both the plasma-dispersion-induced attenuation and radiative recombination processes scale strongly with carrier density. The observed voltage dependence therefore provides strong evidence that the emitted light originates from intrinsic luminescence associated with carrier recombination in the VOA.
\subsection{Central wavelength of parasitic EL}
To unambiguously identify the physical origin of the observed EL, it is essential to determine its spectral position. However, the emission intensity is at the single-photon level, precluding direct spectral analysis using conventional optical spectrum analyzers.

To overcome this limitation, we employ a wavelength-sensitive interferometric technique based on an equal-arm thermo-optic Mach--Zehnder interferometer (MZI), as detailed in~\ref{Measurement Method_Wavelength_Center}. The method exploits the wavelength dependence of the thermo-optic phase shift to infer the center wavelength of an unknown weak light source by comparison with a reference laser of known wavelength.

The center wavelength $\lambda_{\mathrm{VOA}}$ of the VOA emission is determined according to
\begin{equation}
	\lambda_{\mathrm{VOA}}
	= \lambda_{\mathrm{REF}}
	\frac{\bigl(U_{\max,\mathrm{VOA}}^{2} - U_{\min,\mathrm{VOA}}^{2}\bigr)}
	{\bigl(U_{\max,\mathrm{REF}}^{2} -  U_{\min,\mathrm{REF}}^{2}\bigr)} ,
	\label{eq:wavelength_calculation}
\end{equation}
where $\lambda_{\mathrm{REF}}$ is the known wavelength of the reference laser. $U_{\max,\mathrm{VOA}}$ ($ U_{\max,\mathrm{REF}}$) and $U_{\min,\mathrm{VOA}}$ ($ U_{\min,\mathrm{REF}}$) denote the heater voltages corresponding to adjacent interference maxima and minima (separated by a $\pi$ phase shift) for the VOA emission (reference laser).
\begin{figure}[t]
	\centering
	\includegraphics[width=0.85\linewidth]{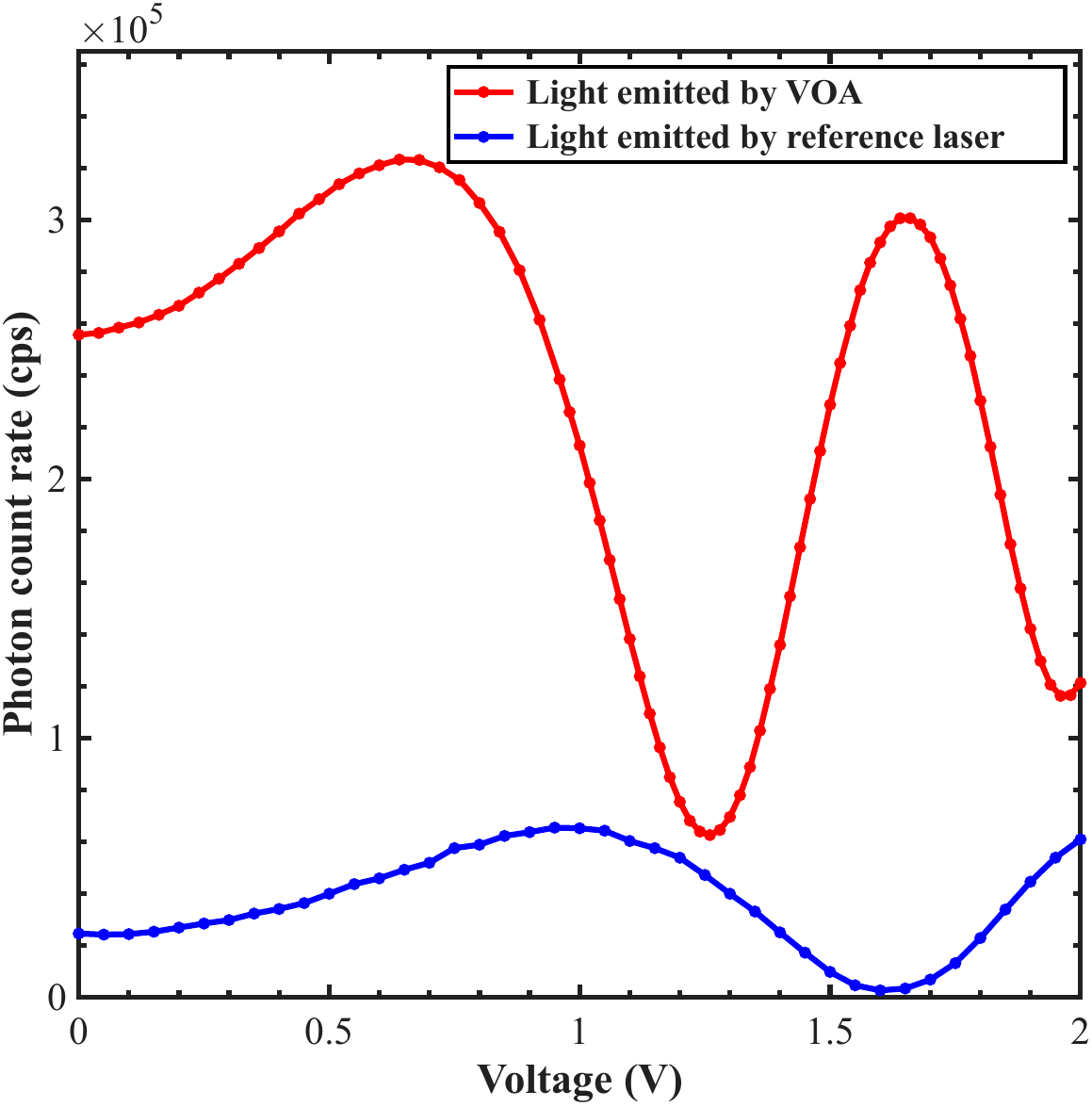}
	\caption{Photon count rates recorded by the $\mathrm{SPD}_{1}$ as a function of the drive voltage applied to the $\mathrm{TOM}_{1}$ in the equal-arm MZI for the VOA emission and a 1550.82~nm reference laser. The count rates exhibit periodic variations with the applied voltage due to interference in the MZI. The filled circles represent the measured data, and the solid lines serve as guides to the eye connecting the data points. The red filled circles correspond to light emitted by the VOA, whereas the blue filled circles correspond to the 1550~nm  reference laser.} \label{CenterWavelength}
\end{figure} 

Experimentally, the upper MZI on the chip [Fig.~\ref{VOA_Experimental_setup}(a)] was used for this measurement. First, an external \SI{1550.82}{\nano\meter} laser attenuated to the single-photon level was injected into the chip, and the $\mathrm{TOM}_1$ voltage was scanned from 0 to \SI{2}{\volt} in steps of \SI{0.01}{\volt}. The resulting interference fringes were recorded by $\mathrm{SPD}_1$. Subsequently, the laser input was removed, the VOA was biased at \SI{1.2}{\volt}, and the same voltage scan was applied to the $\mathrm{TOM}_1$ to record the interference pattern produced by the VOA emission.

As shown in Fig.~\ref{CenterWavelength}, the extracted voltages for the reference laser are $ U_{\max,\mathrm{REF}}=0.95$~V and $ U_{\min,\mathrm{REF}}=1.60$~V, whereas for the VOA emission they are $U_{\max,\mathrm{VOA}}=0.66$~V and $U_{\min,\mathrm{VOA}}=1.26$~V. Substituting these values into Eq.~(\ref{eq:wavelength_calculation}) yields
\begin{equation}
	\lambda_{\mathrm{VOA}} = 1550.82 \cdot \left( \frac{1.26^2 - 0.66^2}{1.60^2 - 0.95^2} \right) \approx 1077.85~\mathrm{nm}.
\end{equation}

The extracted wavelength is in close agreement with the intrinsic radiative recombination wavelength of silicon (1107~nm), providing compelling evidence that the observed emission originates from intrinsic \textit{p--n} junction luminescence rather than from residual guided light or extrinsic optical artifacts.

\section{Security Threats Induced by Parasitic EL in Chip-Based QKD}\label{section4}
In chip-based QKD systems, VOAs are indispensable components for controlling the photon flux. However, as demonstrated in Sec.~\ref{Characterization_VOA}, VOAs can emit parasitic EL that may compromise system security. In this section, we present a detailed analysis of the security threats posed by such parasitic emission, considering two representative transmitter architectures commonly used in integrated QKD platforms.  

\subsection{Case I: VOA Before the Encoder}  

\subsubsection{Security Vulnerability Induced by Parasitic Light}\label{passive_THA_attack}  

\begin{figure}[h]
	\centering
	\includegraphics[width=1\linewidth]{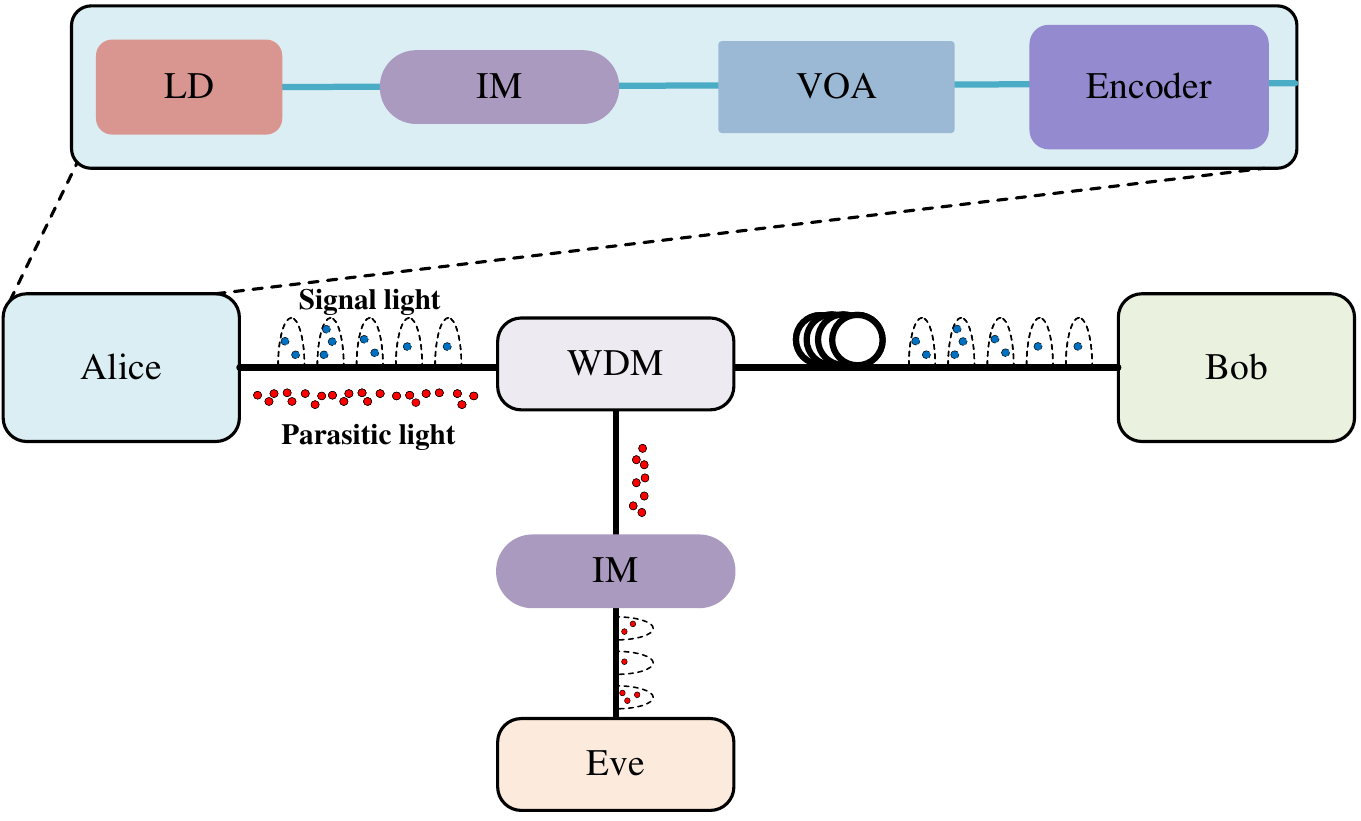}
	\caption{Schematic diagram illustrating the photonic transmitter chip architecture and the corresponding parasitic-light attack. The top panel illustrates the transmitter chip comprising an LD, an IM, an encoder, and a VOA. As the VOA is placed before the encoder, parasitic photons generated by the VOA are encoded with the same information as the signal. The bottom panel depicts Eve's attack strategy: a WDM is used to separate the parasitic light. Subsequently, Eve uses an IM synchronized with the signal pulses to selectively pass the modulated parasitic photons, thereby filtering out the unmodulated ones. LD, laser diode; IM, intensity modulator; VOA, variable optical attenuator; Encoder, polarization encoding module; WDM, wavelength-division multiplexer.}\label{QKD_Structure}
\end{figure}

Figure~\ref{QKD_Structure} depicts a typical chip-based transmitter in which a pulsed laser diode (LD) is followed by an intensity modulator (IM), a VOA, and a polarization encoder. In this \textit{pre-encoder} configuration, the VOA is positioned upstream of the encoding module to reduce polarization-dependent loss~\cite{weiKejin2020,liwei2023,maChaoxuan2016}.  

Under this configuration, parasitic photons generated by the VOA co-propagate with the intended signal photons through the encoder. Consequently, these parasitic photons acquire the same polarization or phase modulation as the signal, creating a potential side channel exploitable by an eavesdropper (Eve).  

Specifically, Eve can separate the parasitic EL photons (e.g., around 1077~nm) from the signal (e.g., 1550~nm) using a wavelength-division multiplexer (WDM). Although the VOA may emit continuously, only photons temporally coincident with the signal pulse carry encoded information. By synchronizing her intensity modulator with Alice's clock, Eve can filter out unmodulated photons, retaining only those carrying key information. These modulated parasitic photons can then be stored in a quantum memory and measured after basis reconciliation, yielding direct access to the raw key.  

This mechanism is analogous to a Trojan-horse attack (THA)~\cite{lucamarini2015}, but with a critical distinction: here, the Trojan photons originate from the intrinsic parasitic emission of the VOA rather than from an injected external source. We therefore refer to this as a \textit{passive Trojan-horse attack (passive THA)}.  

\subsubsection{Quantitative Security Analysis}\label{passive_THA_model}  

To quantify the threat, we model the parasitic emission within the framework of the decoy-state BB84 protocol. Each pulse received by Eve is treated as a coherent state encoding the same quantum information as the signal:  
\begin{equation}
	|\psi_{\phi_i}^i\rangle = |\sqrt{\mu_i}\cos\phi_i\rangle_H \otimes |\sqrt{\mu_i} \sin\phi_i\rangle_V,
	\label{eq:polarization_states}
\end{equation}  

where $\mu_i$ is the mean photon number of the $i$th coherent state, and $\phi_i \in \{0, \pi/2, \pi/4, 3\pi/4\}$ corresponds to the BB84 polarization angles. The total state over $n$ pulses is  
\begin{equation}
	|\Psi\rangle = \bigotimes_{i=1}^n |\psi_{\phi_i}^i\rangle.
	\label{eq:state_total_eve}
\end{equation}  
Based on the experimental characterization (Sec.  \ref{Characterization_VOA}), we assume the leakage is temporally stable, such that $\mu_i = \mu_{\mathrm{Eve}}$ for all $i$. The mean photon number $\mu_{\mathrm{Eve}}$ is related to the measured count rate $C(U)$ and pulse duration $\Delta t$ by:  
\begin{equation}
	\mu_{\mathrm{Eve}} = -\ln \left[ 1 - Q_{\mathrm{Eve}} \right],\quad
	Q_{\mathrm{Eve}} = C(U) \Delta t.
	\label{eq:Eve_mean_photon_number}
\end{equation}  
Here, $Q_{\mathrm{Eve}}$ is the detection probability for Eve under ideal assumptions of unit efficiency and zero dark counts.  

The quantum state sent to Bob can be expressed as the tensor product of the legitimate signal and Eve's parasitic state:  

\begin{equation}
	\begin{aligned}
		|\psi_{z+}\rangle &= |z_{+}\rangle_B \otimes |\psi_0\rangle_E, \\
		|\psi_{z-}\rangle &= |z_{-}\rangle_B \otimes |\psi_{\pi/2}\rangle_E, \\
		|\psi_{x+}\rangle &= |x_{+}\rangle_B \otimes |\psi_{\pi/4}\rangle_E, \\
		|\psi_{x-}\rangle &= |x_{-}\rangle_B \otimes |\psi_{3\pi/4}\rangle_E,
	\end{aligned}
	\label{bb84-states}
\end{equation}  

where $|z_{\pm}\rangle_B$ and $|x_{\pm}\rangle_B$ denote the ideal BB84 states in the $Z$ and $X$ bases, respectively.  

Using this model, in~\ref{gllp_Method}, we apply the GLLP-Koashi approach~\cite{lucamarini2015,GLLP2004,koashi2009} to the states in Eq.~(\ref{bb84-states}) and derive the secure key rate: 
\begin{equation}
	R = p_Z^2 p_1 Y_1 \left[ 1 - h_2(e_X') \right] - p_Z^2 Q_s f h_2(E_s),
	\label{eq:key_rate}
\end{equation}  
where $p_Z$ is the probability of choosing the $Z$ basis, $p_1$ is the single-photon emission probability, $Y_1$ and $e_X'$ are the yield and error rate of single photons under passive THA, $Q_s$ and $E_s$ are the measured gain and QBER of signal pulses, and $f$ is the error correction efficiency.  

The parameter $e_X'$ is given by
\begin{equation}
	\begin{aligned}
		e_X' &= e_X + 4 \Delta' (1 - \Delta') (1 - 2 e_X) \\
		&+ 4 (1 - 2 \Delta') \sqrt{\Delta' (1 - \Delta') e_X (1 - e_X)},
	\end{aligned}
	\label{eq:e_X_prime}
\end{equation}
with
\begin{equation}
	\Delta' = \frac{\Delta}{Y},
	\label{eq:Delta_prime}
\end{equation}
and

\begin{equation}
	\Delta
	= \frac{1}{2}\Biggl[
	1 - e^{-\mu_{\mathrm{Eve}}}
	\Bigl(
	\cosh~\!\Bigl(\frac{\mu_{\mathrm{Eve}}}{\sqrt{2}}\Bigr)
	+ \frac{1}{2}\sinh~\!\Bigl(\frac{\mu_{\mathrm{Eve}}}{\sqrt{2}}\Bigr)
	\Bigr)
	\Biggr],
	\label{eq:Delta}
\end{equation}
where $e_X$ is the error rate of the single-photon component in the $X$ basis, and $Y := \min(Y_Z^1, Y_X^1)$ with $Y_Z^1$ ($Y_X^1$) denoting the single-photon yield in the $Z$ ($X$) basis.

For comparison, we also implement a numerical approach~\cite{liZijian2024,coles2016,liNicky2020,wangwenyuan2022} that directly incorporates the parasitic state in Eq.~\eqref{bb84-states} and optimizes the key rate via a constrained minimization over single-photon states, providing a tighter bound on the secure key rate under realistic leakage conditions (see~\ref{Numerical_Method}).

\subsubsection{Numerical Simulation of Passive THA}

To quantify the impact of passive THA, we simulate a two-decoy-state BB84 protocol under realistic VOA emission conditions. The simulation parameters are based on experimental settings reported in Ref.~\cite{liwei2023} and are summarized in Table~\ref{Parameters}. 

The mean photon number of the leakage, $\mu_\mathrm{Eve}$, depends on the VOA photon count rate $C(U)$ and the signal pulse width $\Delta t$ (Eq.~\eqref{eq:Eve_mean_photon_number}). We consider three representative driving configurations to cover typical operating scenarios: (1) $(U, \Delta t) = (1.4~\mathrm{V}, 200~\mathrm{ps})$: moderate VOA intensity, high repetition rate ($\sim2.5~\mathrm{GHz}$),
(2) $(U, \Delta t) = (1.4~\mathrm{V}, 1.6~\mathrm{ns})$: moderate intensity, lower repetition rate ($\sim625~\mathrm{MHz}$),
(3)$(U, \Delta t) = (2.0~\mathrm{V}, 1.6~\mathrm{ns})$: maximum VOA intensity, lower repetition rate ($\sim625~\mathrm{MHz}$). 

These scenarios cover both typical and extreme operating conditions. The measured $C(U)$ values are listed in Table~\ref{Parameters}. Using these values, We obtain $\mu_{\mathrm{Eve}}^{(1)}=0.0048$, $\mu_{\mathrm{Eve}}^{(2)}=0.0388$, and $\mu_{\mathrm{Eve}}^{(3)}=0.0977$. Notably, higher VOA drive voltages and longer pulse widths increase parasitic EL, resulting in higher information leakage.

\begin{figure}[h!]
	\centering
	\includegraphics[width=0.85\linewidth]{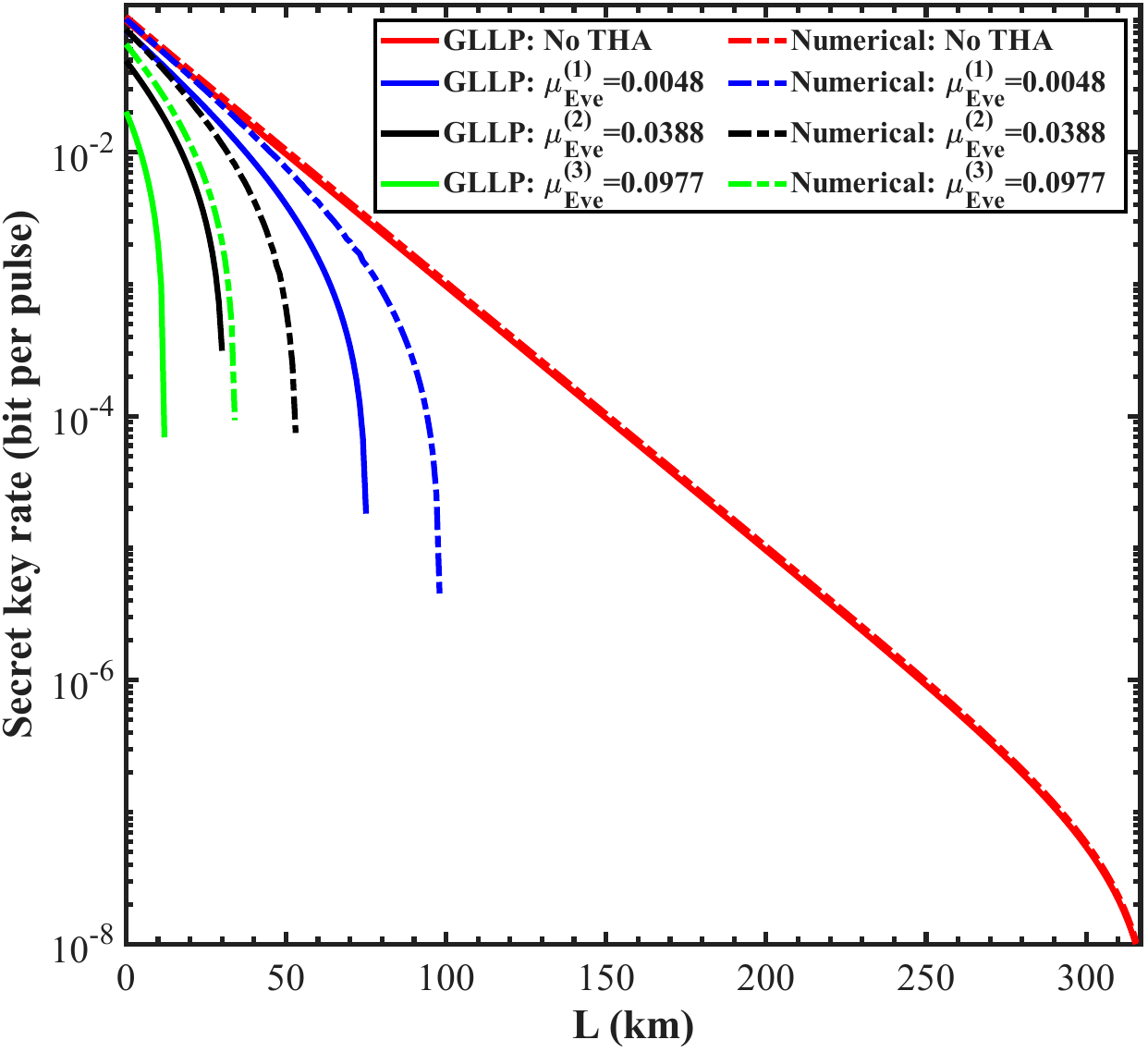}
	\caption{Simulated secret key rate versus transmission distance for the two-decoy-state BB84 protocol under passive THA. Solid (dashed) curves correspond to the conventional GLLP analysis (numerical method). Red curves show the ideal case without THA. Blue, black, and green curves correspond to the three scenarios with $\mu_{\mathrm{Eve}}^{(1)}=0.0048$, $\mu_{\mathrm{Eve}}^{(2)}=0.0388$, and $\mu_{\mathrm{Eve}}^{(3)}=0.0977$, respectively, where $\mu_{\mathrm{Eve}}^{(1)}$, $\mu_{\mathrm{Eve}}^{(2)}$, and $\mu_{\mathrm{Eve}}^{(3)}$ are associated with the driving configurations $(U,\Delta t)=(1.4~\mathrm{V},200~\mathrm{ps})$, $(1.4~\mathrm{V},1.6~\mathrm{ns})$, and $(2.0~\mathrm{V},1.6~\mathrm{ns})$, respectively. For all curves, the signal intensity is set to $s=0.48$, with two decoy intensities $\nu=0.02$ and $w=0.001$.} 
	\label{keyrate1}
\end{figure}

The simulation results are presented in Fig.~\ref{keyrate1}. We observe that: (1) The numerical method consistently yields higher key rates and longer maximum distances than the conventional GLLP analysis, reflecting its advantage in providing tighter bounds on the information leakage under device imperfections through the full utilization of observed statistics and source constraints. (2) Passive THA substantially degrades system performance. In the worst-case scenario $(2.0~\mathrm{V}, 1.6~\mathrm{ns})$, the maximum transmission distance is reduced from $320~\mathrm{km}$ (ideal case) to $19~\mathrm{km}$, with the key rate falling by more than an order of magnitude.
(3) Increasing either the VOA drive voltage $U$ or the pulse width $\Delta t$ increases leakage ($\mu_\mathrm{Eve} \propto C(U) \Delta t$), requiring stronger privacy amplification. For example, at $\Delta t = 1.6~\mathrm{ns}$, increasing $U$ from $1.4~\mathrm{V}$ to $2.0~\mathrm{V}$ reduces the maximum distance by more than $15~\mathrm{km}$ and suppresses the key rate by roughly an order of magnitude. Likewise, extending the pulse width from $200~\mathrm{ps}$ to $1.6~\mathrm{ns}$ at $U=1.4~\mathrm{V}$ decreases the reach by over $40~\mathrm{km}$, accompanied by a significant drop in key rate.

These results highlight that even moderate VOA parasitic emissions can seriously compromise the security of chip-based QKD systems. Careful optimization of VOA drive conditions and pulse widths is therefore critical for practical implementations.
\begin{table}[t!] 
	\setlength\tabcolsep{3pt} 
	\centering
	\caption{Simulation parameters for the QKD system and measured photon count rates $C(U)$ of VOA parasitic light at different applied voltages $U$. The system parameters are adopted from Ref.~\cite{liwei2023}: $\eta_{\text{Bob}}$ is the detection efficiency, $e_d$ the misalignment error, $Y_0$ the dark count rate, and $f_{e}$ the error correction inefficiency. The photon count rates $C(U)$ used in the simulations are obtained from experimental measurements.}
	\begin{tabular}{cccccc}
		\hline\hline\
		$e_d$   & $Y_{0}$          & $\eta_{Bob}$ & $f_{e}$  & $C(1.4V)$ & $C(2V)$\\ \hline
		0.0061  & $2\times10^{-8}$ & 0.78         & 1.2   &$2.38\times10^{7}$ & $5.82\times10^{7}$\\ \hline\hline
	\end{tabular}\label{Parameters}
\end{table}

\subsection{Case II: VOA After the Encoder}\label{section6}
\subsubsection{Parasitic Noise Effects}
\begin{figure}[h]
	\centering
	\includegraphics[width=1\linewidth]{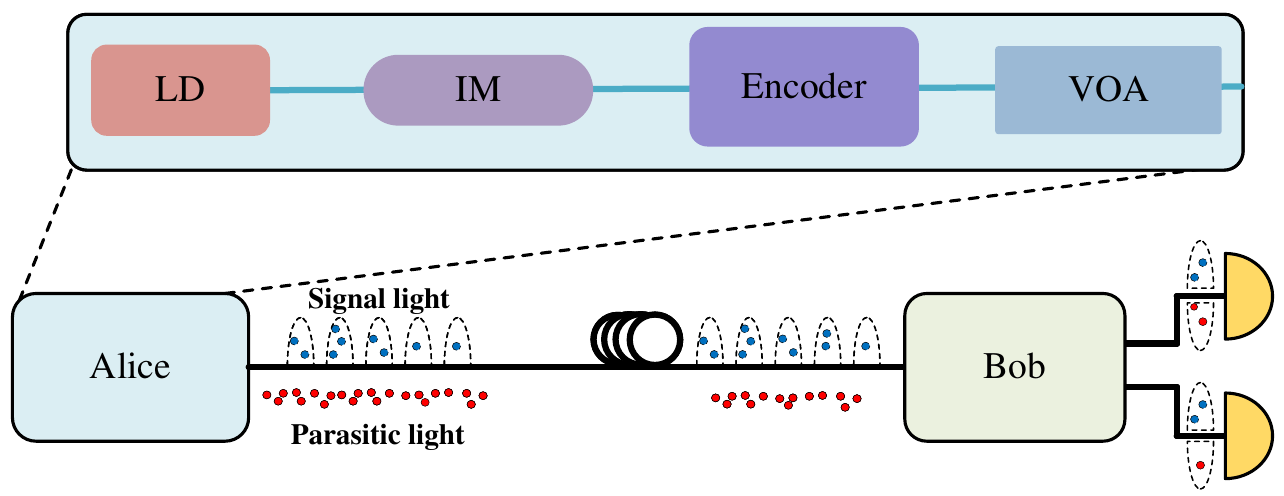}
	\caption{Schematic diagram of a chip-based QKD transmitter with a post-encoder VOA and its impact on QKD. The VOA generates parasitic photons that bypass the encoder and co-propagate with the signal. These unmodulated photons can induce additional detector clicks at Bob's side, thereby increasing the observed gain. LD: laser diode; IM: intensity modulator; Encoder: quantum state encoder; VOA: variable optical attenuator.}
	\label{QKD_Structure2}
\end{figure}

Figure~\ref{QKD_Structure2} illustrates a typical chip-based QKD transmitter configuration in which the VOA is positioned \emph{after} the quantum state encoder, in contrast to the pre-encoder VOA architecture discussed in Sec.~\ref{passive_THA_attack}. This post-encoder arrangement is commonly employed in phase-encoding QKD systems~\cite{bacco2017}. In this configuration, parasitic photons generated by the VOA are directly injected into the quantum channel alongside the encoded signal photons. Importantly, these parasitic photons bypass the encoding module and therefore remain unmodulated, carrying no quantum information.

While parasitic emissions may occur outside the temporal duration of the signal pulses, Bob's detection system only records arrival events within predefined signal time windows. Consequently, the parasitic photons appear as synchronized pulse trains at the receiver, indistinguishable in timing from the legitimate signal.

Although unencoded, these parasitic photons act as an additional noise source. Even when Bob's detectors are optimized for the signal wavelength (e.g., 1550~nm), they typically exhibit non-negligible sensitivity at other wavelengths, including the VOA emission band (e.g., around 1077~nm). As a result, parasitic photons can generate spurious detection events, artificially increasing the observed gain. If such contributions are not separately characterized, they bias the parameter estimation used in decoy-state analysis, potentially leading to an overestimation of the secret key rate. We refer to this loophole as the \emph{dual-source flaw}, which consists of the intended signal source and a parasitic noise source.

\subsubsection{Security Modeling of the dual-source flaw}\label{dual-source flaw Model}
In the dual-source flaw, the system is modeled as two independent weak coherent sources (WCSs): the primary signal source with intensity $\gamma$ (selected from the decoy-state intensities, e.g., $\gamma \in \{s, \nu, \omega\}$) and a parasitic noise source with intensity $\mu_{\scriptscriptstyle \mathrm{EL}}$ arising from the VOA parasitic EL. Due to the stable VOA operating voltage, $\mu_{\scriptscriptstyle \mathrm{EL}}$ can be considered constant over time.

The mean photon number of the parasitic source is determined by an idealized calibration using a single-photon detector with unit efficiency ($\eta_{\mathrm{Bob}} = 1$) and zero dark counts ($Y_{0,\mathrm{Bob}}=0$). For a temporal pulse width $\Delta t$, the observed gain is $Q_{\scriptscriptstyle \mathrm{EL}} = C(U) \Delta t$, leading to
\begin{equation} \label{eq:mu_leak_calculation}
	\mu_{\scriptscriptstyle \mathrm{EL}} = -\ln\left( 1 - C(U) \cdot \Delta t \right),
\end{equation}
where $C(U)$ is the experimentally measured photon count rate of the VOA parasitic emission at driving voltage $U$.

Both the signal and parasitic sources are modeled as Poissonian:
\begin{align}
	\rho_{\gamma} &= \sum_{n=0}^{\infty} \frac{\gamma^n}{n!} e^{-\gamma} |n\rangle \langle n|, \\
	\rho_{\mu_{\scriptscriptstyle \mathrm{EL}}} &= \sum_{n=0}^{\infty} \frac{\mu_{\scriptscriptstyle \mathrm{EL}}^n}{n!} e^{-\mu_{\scriptscriptstyle \mathrm{EL}}} |n\rangle \langle n|.
\end{align}

Under the assumption of independent sources and a standard optical-fiber channel model with threshold detectors~\cite{MaXiongfeng2005}, the overall gain $Q_{\gamma,\mu_{\scriptscriptstyle \mathrm{EL}}}$ and the corresponding QBER $E_{\gamma,\mu_{\scriptscriptstyle \mathrm{EL}}}$ are derived. Detailed derivations are provided in~\ref{Q_EQ}, and the resulting \begin{equation}
	Q_{\gamma,\mu_{\scriptscriptstyle \mathrm{EL}}}
	= 1 - (1-Y_0)\, e^{-\gamma\eta}\, e^{-\mu_{\scriptscriptstyle \mathrm{EL}}\eta'},
	\label{Q_all}
\end{equation}  
\begin{equation}
	\begin{aligned}
		E_{\gamma,\mu_{\scriptscriptstyle \mathrm{EL}}}\,Q_{\gamma,\mu_{\scriptscriptstyle \mathrm{EL}}}
		&= Y_0 e_0 + e_d \bigl(1 - e^{-\gamma \eta}\bigr) + e_0 \bigl(1 - e^{-\mu_{\scriptscriptstyle \mathrm{EL}}\eta'}\bigr) \\
		& - Y_0 e_0 e_d \bigl(1 - e^{-\gamma \eta}\bigr) - Y_0 e_0^2 \bigl(1 - e^{-\mu_{\scriptscriptstyle \mathrm{EL}}\eta'}\bigr) \\
		& - e_d e_0 \bigl(1 - e^{-\gamma \eta}\bigr)\bigl(1 - e^{-\mu_{\scriptscriptstyle \mathrm{EL}}\eta'}\bigr) \\
		& + Y_0 e_0^2 e_d \bigl(1 - e^{-\gamma \eta}\bigr)\bigl(1 - e^{-\mu_{\scriptscriptstyle \mathrm{EL}}\eta'}\bigr).
	\end{aligned}
	\label{EQ_all}
\end{equation}
where $Y_0$ is the background rate, $\eta$ ($\eta'$) is the system transmittance for the signal (parasitic light), $e_d$ is the intrinsic misalignment error, and $e_0$ is the error probability of background counts.

Based on the above model, we perform a security analysis of the dual-source flaw, as detailed in~\ref{multi-light source_key_rate}. The resulting secret key rate is given by
\begin{equation} \label{eq:DSS_key-rate-standard}
	R \geq q \left\{ Q_1^L [1 - h_2(e_1^U)] - Q_{s,\mu_{\scriptscriptstyle \mathrm{EL}}} f_{\text{EC}} h_2(E_{s,\mu_{\scriptscriptstyle \mathrm{EL}}}) \right\},
\end{equation}
where $q$ is the protocol efficiency (e.g., $q=1/2$ for BB84), $Q_1^L$ and $e_1^U$ denote the lower bound on single-photon yield and upper bound on single-photon error rate, respectively, and $f_{\text{EC}}$ is the error correction efficiency.

\subsubsection{Simulation of Parasitic-Light Effects} \label{section6B}

To quantify the impact of VOA parasitic emission on QKD security, we perform numerical simulations using a two-decoy-state BB84 protocol. The goal is to compare the achievable secure key rate under two scenarios: the \emph{no parasitic light case} (where parasitic photons are ideally filtered out) and the \emph{parasitic-light case} (where parasitic photons co-propagate with the signal). 

We adopt the same experimental parameters as those used in the passive THA simulations (Table~\ref{Parameters}). The parasitic-light mean photon number $\mu_{\scriptscriptstyle \mathrm{EL}}$ is calculated according to Eq.~\eqref{eq:mu_leak_calculation}, using the measured photon count rate $C(U)$ and pulse width $\Delta t$. Three representative driving configurations are considered: $(U, \Delta t)=(1.4~\text{V}, 200~\text{ps})$, $(1.4~\text{V}, 1.6~\text{ns})$, and $(2.0~\text{V}, 1.6~\text{ns})$, which correspond to $\mu_{\scriptscriptstyle \mathrm{EL}}^{(1)}=0.0048$, $\mu_{\scriptscriptstyle \mathrm{EL}}^{(2)}=0.0388$, and $\mu_{\scriptscriptstyle \mathrm{EL}}^{(3)}=0.0977$, respectively.

The parasitic light has a wavelength of $1077~\mathrm{nm}$, distinct from the signal wavelength ($1550~\mathrm{nm}$), resulting in different fiber attenuation and detection efficiency. We set the fiber attenuation and Bob's overall detection efficiency for the parasitic light to: $
\alpha' = 0.8~\mathrm{dB/km}, \quad \eta'_{\mathrm{Bob}} = 25\%.$
\begin{figure}[h!]
	\centering
	\includegraphics[width=0.85\linewidth]{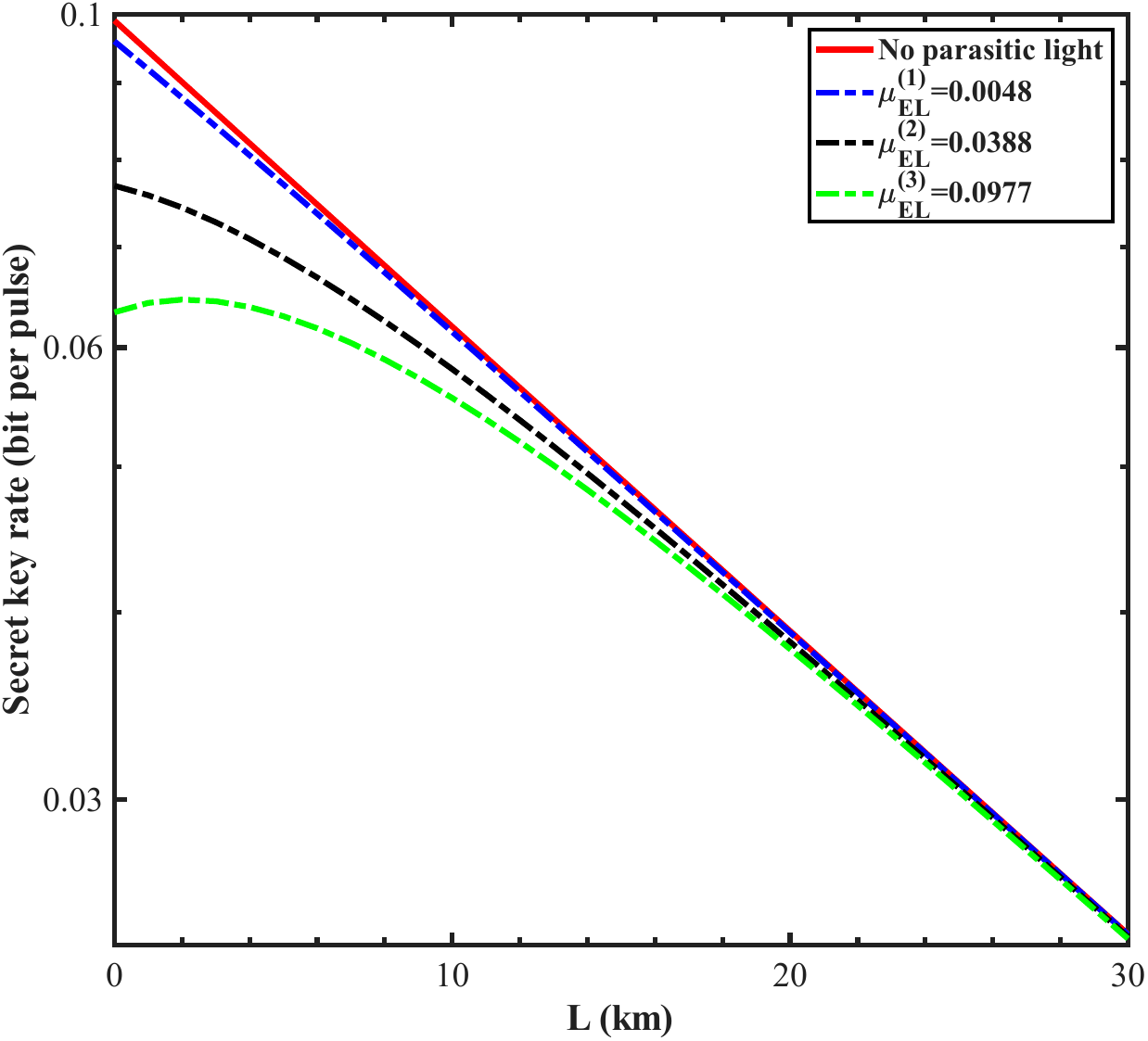}
	\caption{Secure key rate as a function of transmission distance considering the impact of parasitic light. The red solid line represents the \emph{no parasitic light case}. The dashed lines correspond to the \emph{parasitic-light cases} with mean photon numbers of $\mu_{\scriptscriptstyle \mathrm{EL}}^{(1)}=0.0048$ (blue), $\mu_{\scriptscriptstyle \mathrm{EL}}^{(2)}=0.0388$ (black), and $\mu_{\scriptscriptstyle \mathrm{EL}}^{(3)}=0.0977$ (green). These values are obtained under driving configurations $(U, \Delta t)$ of $(1.4\,\text{V}, 200\,\text{ps})$, $(1.4\,\text{V}, 1.6\,\text{ns})$, and $(2.0\,\text{V}, 1.6\,\text{ns})$, respectively. For all curves, the signal intensity is set to $s=0.48$, with two decoy intensities $\nu=0.02$ and $w=0.001$.}\label{keyrate_VOA_afte1}
\end{figure}

The simulation results are presented in Fig.~\ref{keyrate_VOA_afte1}. Key observations include: (1) At short transmission distances ($< 10~\mathrm{km}$), parasitic light significantly reduces the secure key rate. The worst-case scenario $(2.0~\text{V}, 1.6~\text{ns})$ causes approximately 50\% reduction relative to the ideal case.
(2) The impact diminishes rapidly with increasing distance due to higher fiber attenuation for the parasitic wavelength. For distances $\gtrsim 15~\mathrm{km}$, the contribution of parasitic photons at Bob becomes negligible, and the secure key rate approaches the ideal value.
(3) The effect of parasitic light is more pronounced at higher VOA driving voltages and longer pulse widths, consistent with the increased mean photon number $\mu_{\scriptscriptstyle \mathrm{EL}}$.

These results emphasize the importance of proper VOA design, parasitic-light mitigation (e.g., spectral filtering), and calibration protocols to ensure secure operation of chip-based QKD systems.  

\section{Discussion and Conclusions}\label{section7}

In this work, we have identified and characterized a previously overlooked source of noise and potential vulnerability in chip-based quantum key distribution (QKD) systems: parasitic light emission from integrated VOAs. Through a combination of experimental measurements, physical modeling, and security analysis, we have demonstrated that VOAs, while performing their intended optical attenuation function, can emit weak luminescence. Although the emitted light is orders of magnitude weaker than the signal, our analysis shows that it can still induce measurable detection events at the receiver, introducing a dual-source flaw that biases key parameter estimation in decoy-state protocols.

Our methodology integrates several layers of investigation. First, we experimentally quantified the parasitic emission intensity and temporal characteristics of the VOAs under various operating conditions. Second, we developed a theoretical model treating the parasitic light as an independent weak coherent source, allowing us to compute its contribution to the overall gain and quantum bit error rate (QBER) at the receiver. Third, using this model, we performed simulations of standard two-decoy-state BB84 QKD to evaluate the resulting impact on secure key rates under realistic system parameters. This combination of experimental characterization, theoretical modeling, and protocol-level simulation provides a rigorous framework for assessing the security implications of device-level imperfections in integrated photonic QKD.

The results reveal that even weak parasitic light can substantially affect key rates, particularly at short transmission distances or under operating conditions that maximize VOA emission. Importantly, this effect is distinct from previously studied active side-channel attacks: it arises from an intrinsic physical property of the device rather than deliberate manipulation by an adversary. This finding highlights a broader category of hardware vulnerabilities in integrated QKD systems, wherein parasitic physical effects inherent to semiconductor-based components, such as \textit{p-n} junction devices, can compromise security if unaccounted for.

Looking forward, our study suggests several directions for future work. Systematic characterization of parasitic emissions should be extended to other integrated photonic devices, such as phase modulators and on-chip detectors, which may exhibit similar unintended luminescence. Moreover, security analyses should encompass a wider variety of QKD protocols, including measurement-device-independent QKD~\cite{Hoi-Kwong-lo2012}  and twin-field QKD~\cite{lucamarini2018}, to determine the generality of these vulnerabilities. Finally, strategies for mitigation-ranging from improved device design and spectral filtering to calibration protocols that explicitly account for parasitic sources--should be developed to ensure the reliable deployment of high-rate, chip-integrated QKD systems.

In summary, our work establishes a new paradigm for understanding and mitigating hardware-induced vulnerabilities in integrated QKD. By revealing the potential security impact of parasitic VOA emissions, we provide a foundation for both improving device design and informing comprehensive security proofs for next-generation photonic quantum communication technologies.

\vspace{1cm}
\noindent 
\textbf{Acknowledgments}~~We are profoundly grateful to Prof. Anqi Huang (National University of Defense Technology) and Prof. Feihu Xu (University of Science and Technology of China) for their insightful discussions and valuable suggestions.

\vspace{0.5cm}
\noindent 
\textbf{Funding}~~This study was supported by the National Natural Science Foundation of China (62171144, 62031024); Guangxi Science Foundation (2025GXNSFAA069137, GXR-1BGQ2424005); Innovation Project of Guangxi Graduate Education (YCBZ2025064).

\vspace{0.5cm}
\noindent 
\textbf{Author Contributions}~~K.W. conceived the original idea and directed the research. Z.L. performed the experiments and analyzed the data. C.X. assisted in the experiments. X.H. and X.X. provide the tested chip. Y.D., X.L. and T.L. contributed to helpful discussions and the interpretation of the results. All authors reviewed and revised the manuscript.

\vspace{0.5cm}
\noindent 
\textbf{Competing Interests}~~The authors declare no competing interests.

\vspace{0.5cm}
\noindent 
\textbf{Data availability}~~The data that support the findings of this study are available from the corresponding author upon reasonable request.

\appendix

\section{Central Wavelength Measurement Method}
\label{Measurement Method_Wavelength_Center}

To unambiguously identify the physical origin of the parasitic emission from the on-chip VOA, we develop a wavelength-measurement method capable of operating at the single-photon level. The method exploits the wavelength dependence of thermo-optic phase modulation in an equal-arm Mach--Zehnder interferometer (MZI) and does not require prior calibration of device-specific parameters.

\begin{figure}[h]
	\centering
	\includegraphics[width=1\linewidth]{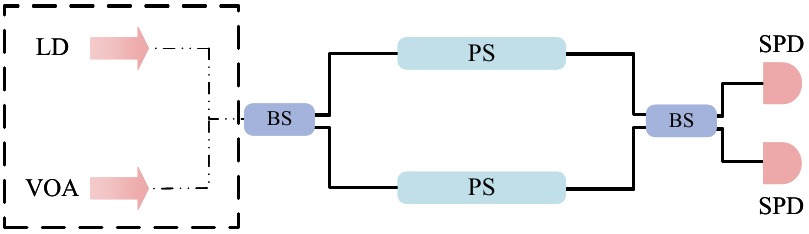}
	\caption{Schematic of the central-wavelength measurement setup based on an equal-arm Mach--Zehnder interferometer (MZI). LD: attenuated reference laser; VOA: on-chip variable optical attenuator; BS: beam splitter; PS: thermo-optic phase shifter; SPD: single-photon detector.}
	\label{Wavelength_setup}
\end{figure}

The measurement configuration is illustrated in Fig.~\ref{Wavelength_setup}. The setup consists of an on-chip equal-arm MZI formed by two beam splitters (BSs) and a thermo-optic phase shifter (PS) placed in one interferometer arm. A single-photon detector (SPD) is connected to one output port of the MZI. The method proceeds by comparing the interference response of the unknown emission to that of a reference light source with a known central wavelength.

First, light from a laser diode (LD) with known center wavelength $\lambda_{\mathrm{REF}}$ is attenuated to the single-photon level and injected into the MZI. A voltage $U$ is applied to the PS, inducing a thermo-optic phase shift in the heated arm. By scanning $U$ and recording the photon count rate at the SPD, an interference fringe pattern is obtained as a function of the applied voltage. The same procedure is subsequently repeated using the parasitic emission from the VOA, whose central wavelength $\lambda_{\mathrm{VOA}}$ is unknown.

The thermo-optic phase accumulated in the PS of length $L$ for light at wavelength $\lambda$ can be written as
\begin{equation}
	\phi_{\mathrm{PS}}(U)
	= \frac{2\pi L}{\lambda}
	\bigl(n_{\lambda} + \eta P(U)\bigr),
	\label{eq:phase_general}
\end{equation}
where $n_{\lambda}$ is the effective refractive index of the unheated waveguide, $\eta$ is the effective thermo-optic coefficient relating refractive-index change to dissipated electrical power, and $P(U)$ is the heater power. The electrical power depends on the applied voltage as
\begin{equation}
	P(U) = \frac{U^2}{R},
	\label{eq:power_voltage_appendix}
\end{equation}
with $R$ denoting the heater resistance. Over the spectral range relevant to this work, $L$, $R$, and $\eta$ are assumed to be wavelength independent.

When the reference light at wavelength $\lambda_{\mathrm{REF}}$ propagates through the MZI, the total optical phase in the heated (upper) arm is
\begin{equation}
	\phi_{\mathrm{up}}(U)
	= \phi_{\mathrm{up}}^{(0)}
	+ \frac{2\pi L \eta}{\lambda_{\mathrm{REF}} R} U^2,
	\label{eq:phi_up_ref}
\end{equation}
where $\phi_{\mathrm{up}}^{(0)}$ is the static phase in the absence of heating. The lower arm remains unheated, with a constant phase $\phi_{\mathrm{down}}^{(0)}$. The resulting phase difference between the two arms is therefore
\begin{equation}
	\Delta\phi(U)
	= \Delta\phi_0
	+ \frac{2\pi L \eta}{\lambda_{\mathrm{REF}} R} U^2,
	\label{eq:phase_difference_ref}
\end{equation}
where $\Delta\phi_0 = \phi_{\mathrm{up}}^{(0)} - \phi_{\mathrm{down}}^{(0)}$ represents the static phase imbalance of the interferometer.

As the voltage is scanned, the photon count rate at the detector oscillates between interference maxima and minima. Let $ U_{\min,\mathrm{REF}}$ and $ U_{\max,\mathrm{REF}}$  denote the voltages corresponding to two adjacent extrema separated by a $\pi$ phase shift, such that
\begin{equation}
	\Delta\phi( U_{\max,\mathrm{REF}}) - \Delta\phi( U_{\min,\mathrm{REF}}) = \pi.
\end{equation}
Substituting Eq.~\eqref{eq:phase_difference_ref} and eliminating the static phase term yields
\begin{equation}
	\frac{2\pi L \eta}{\lambda_{\mathrm{REF}} R}
	\left( U_{\max,\mathrm{REF}}^2 -  U_{\min,\mathrm{REF}}^2\right) = \pi,
\end{equation}
which can be rearranged as
\begin{equation}
	\lambda_{\mathrm{REF}}
	= \frac{2 L \eta}{R}
	\left( U_{\max,\mathrm{REF}}^2 -  U_{\min,\mathrm{REF}}^2\right) .
	\label{eq:lambda_ref_appendix}
\end{equation}

An analogous relation holds when the reference laser is replaced by the VOA emission with unknown center wavelength $\lambda_{\mathrm{VOA}}$:
\begin{equation}
	\lambda_{\mathrm{VOA}}
	= \frac{2 L \eta}{R}
	\left(U_{\max,\mathrm{VOA}}^2 - U_{\min,\mathrm{VOA}}^2\right).
	\label{eq:lambda_voa_appendix}
\end{equation}
Taking the ratio of Eqs.~\eqref{eq:lambda_voa_appendix} and~\eqref{eq:lambda_ref_appendix} eliminates all device-dependent parameters, yielding
\begin{equation}
	\lambda_{\mathrm{VOA}}
	= \lambda_{\mathrm{REF}}
	\frac{\bigl(U_{\max,\mathrm{VOA}}^{2} - U_{\min,\mathrm{VOA}}^{2}\bigr)}
	{\bigl( U_{\max,\mathrm{REF}}^{2} -  U_{\min,\mathrm{REF}}^{2}\bigr)} .
	\label{appendix_eq:wavelength_calculation}
\end{equation}
This relation enables direct extraction of the central wavelength of the VOA emission from measured interference fringes, without requiring knowledge of the interferometer geometry, heater resistance, or thermo-optic coefficient. The method is therefore well suited for characterizing extremely weak on-chip light emission at the single-photon level.

\section{Verification of the \textit{p--n} Junction Structure}
\label{Confirmation_PN}

To elucidate the physical origin of the parasitic EL observed in the VOA, we experimentally verify whether the device incorporates an intrinsic \textit{p--n} junction. The presence of such a junction is a prerequisite for carrier injection and radiative recombination under forward bias. A standard and unambiguous diagnostic is provided by the current--voltage ($I$--$V$) characteristic, as a \textit{p--n} junction exhibits a distinct nonlinear response governed by diode transport physics.

According to the Shockley diode model, the forward-bias current of an ideal \textit{p--n} junction follows an exponential dependence on the applied voltage,
\begin{equation}
	I \propto \exp\!\left(\frac{qV}{\beta k T}\right),
\end{equation}
where $q$ is the elementary charge, $k$ is the Boltzmann constant, $T$ is the absolute temperature, and $\beta$ is the ideality factor. The parameter $\beta $ captures the dominant carrier transport mechanism and typically varies with bias voltage in practical devices.   

\begin{figure}[h]
	\centering
	\includegraphics[width=0.85\linewidth]{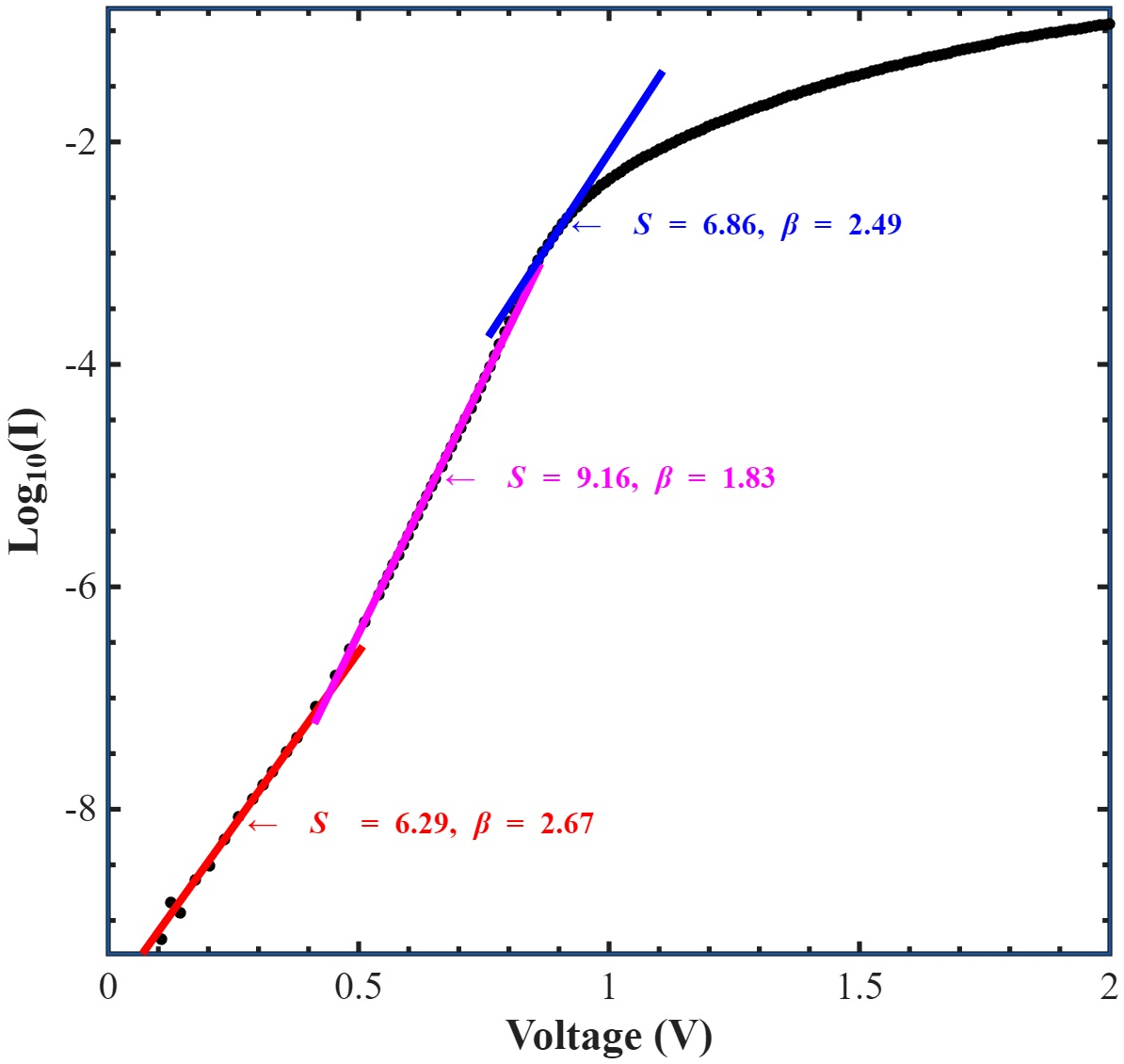}
	\caption{Semi-logarithmic current--voltage characteristic of the VOA under forward bias. The current is plotted on a logarithmic scale, and the voltage on a linear scale. Black symbols represent experimental data. Solid lines indicate linear fits in three bias regions: 0--0.45\,V (red), 0.50--0.80\,V (magenta), and 0.85--0.90\,V (blue). The extracted slopes are used to determine the ideality factor $n$ in each regime.}
	\label{V_logI}
\end{figure}

Figure~\ref{V_logI} shows the measured forward-bias $I$--$V$ characteristic of the VOA plotted on a semi-logarithmic scale. The data exhibit multiple quasi-linear regions, indicating exponential current-voltage relations with different effective ideality factors. Such behavior is a hallmark of \textit{p--n} junction transport and reflects transitions between distinct conduction regimes as the bias increases.

At low forward bias, carrier injection is weak and the current is dominated by recombination processes within the space-charge region, leading to an ideality factor close to $\beta  \approx 2$. With increasing bias, diffusion current becomes dominant and the ideality factor approaches the diffusion-limited value $\beta  \approx 1$. At still higher bias, high-level injection effects emerge, causing the ideality factor to increase again toward $\beta  \approx 2$. Finally, at sufficiently large voltages, series resistance limits the current, and the $I$--$V$ curve deviates from exponential behavior.

To quantitatively extract the ideality factor, we analyze the slope of the semi-logarithmic $I$--$V$ curve following the method of Ref.~\cite{seiferth1994}. Writing the Shockley relation in base-10 logarithmic form,
\begin{equation}
	\log_{10}(I) = \frac{q \log_{10}(e)}{\beta k T} V,
\end{equation}
the slope $S = d\log_{10}(I)/dV$ is given by
\begin{equation}
	S = \frac{q \log_{10}(e)}{\beta k T}.
\end{equation}
The ideality factor can therefore be expressed as
\begin{equation}
	\beta = \frac{q \log_{10}(e)}{S k T}
	\approx \frac{q}{2.3\, S k T},
	\label{n_eq}
\end{equation}
where the numerical factor arises from the conversion between natural and base-10 logarithms.

By performing linear fits on the data within the three bias regions indicated in Fig.~\ref{V_logI}, we extract the corresponding slopes and ideality factors. In the low-bias region (0--0.45\,V), the extracted value $\beta \approx 2.6$ is consistent with recombination-dominated transport. In the intermediate-bias region (0.50--0.80\,V), $\beta $ decreases to approximately 1.8, approaching the diffusion-limited regime. At higher bias (0.85--0.90\,V), $\beta $ increases again to $\sim 2.5$, signaling the onset of high-level injection. Beyond 0.90\,V, the curve progressively deviates from exponential behavior, indicating the influence of series resistance.

Although the extracted ideality factors deviate slightly from their ideal integer values, such deviations are expected in practical integrated devices due to effects including contact resistance and material inhomogeneity. Importantly, the observed evolution of $\beta$ with bias---from recombination-dominated to diffusion-dominated and back to high-injection behavior---constitutes a clear and internally consistent signature of a \textit{p--n} junction. This provides direct experimental confirmation that the VOA incorporates an active \textit{p--n} junction capable of supporting electrically driven EL.

\section{Key Rate Analysis under Passive THA via the GLLP Method} \label{gllp_Method}  
In this appendix, we analyze the polarization-encoded BB84 protocol within the GLLP--Koashi security framework~\cite{GLLP2004,koashi2009}. In the presence of a passive THA, the quantum state emitted by Alice can be written as
\begin{equation}
	\begin{aligned}
		\ket{\psi_{z+}}_{BE} &= \ket{z_{+}}_{B} \otimes \ket{\psi_{0}}_{E}, \\
		\ket{\psi_{z-}}_{BE} &= \ket{z_{-}}_{B} \otimes \ket{\psi_{\pi/2}}_{E}, \\
		\ket{\psi_{x+}}_{BE} &= \ket{x_{+}}_{B} \otimes \ket{\psi_{\pi/4}}_{E}, \\
		\ket{\psi_{x-}}_{BE} &= \ket{x_{-}}_{B} \otimes \ket{\psi_{3\pi/4}}_{E},
	\end{aligned}
	\label{appendix_bb84-states}
\end{equation}
where
\begin{equation}
	\begin{aligned}
		\ket{\psi_{\phi}}_{E}
		&=
		\Bigl\lvert \sqrt{\mu_{\mathrm{Eve}}}\cos\phi \Bigr\rangle_{H}
		\otimes
		\Bigl\lvert \sqrt{\mu_{\mathrm{Eve}}}\sin\phi \Bigr\rangle_{V},\\[4pt]
		\ket{z_{+}}_{B} &= \ket{H}, \qquad
		\ket{z_{-}}_{B} = \ket{V},\\[4pt]
		\ket{x_{+}}_{B} &= \frac{\ket{H}+\ket{V}}{\sqrt{2}}, \qquad
		\ket{x_{-}}_{B} = \frac{\ket{H}-\ket{V}}{\sqrt{2}}.
	\end{aligned}
\end{equation}

We use an entanglement-based description instead of the prepare-and-measure scheme. The $Z$-basis states in Eq.~\eqref{appendix_bb84-states} can be prepared by Alice by measuring in the basis $\{\ket{z_+}_A, \ket{z_-}_A\}$ the following entangled state:
\begin{equation}
	\ket{\Psi_{Z}}
	=
	\frac{
		\ket{z_{+}}_{A}\ket{\psi_{z+}}_{BE}
		+
		\ket{z_{-}}_{A}\ket{\psi_{z-}}_{BE}
	}{\sqrt{2}}.
	\label{eq:PsiZ_app}
\end{equation}
Similarly, the $X$-basis states in Eq.~\eqref{appendix_bb84-states} can be prepared by Alice by measuring subsystem $A$ in the basis $\{\ket{x_+}_A, \ket{x_-}_A\}$ of the following entangled state:
\begin{equation}
	\ket{\Psi_{X}}
	=
	\frac{
		\ket{x_{+}}_{A}\ket{\psi_{x+}}_{BE}
		+
		\ket{x_{-}}_{A}\ket{\psi_{x-}}_{BE}
	}{\sqrt{2}}.
	\label{eq:PsiX_app}
\end{equation}

To quantify the basis dependence of Alice's source, we adopt the ``quantum coin'' technique of
Refs.~\cite{GLLP2004} by introducing an additional qubit $C$ (the quantum coin), also held by Alice, which encodes the choice of preparation basis. The state is defined as:
\begin{equation}
	\ket{\Phi}
	=
	\frac{
		\ket{z_{+}}_{C}\ket{\Psi_{Z}}
		+
		\ket{z_{-}}_{C}\ket{\Psi_{X}}
	}{\sqrt{2}},
	\label{eq:Phi_CABE}
\end{equation} 

Following Ref.~\cite{GLLP2004}, the basis dependence of Alice's source is quantified by the probability that the two entangled states $\ket{\Psi_{Z}}$ and $\ket{\Psi_{X}}$ are not identical. In the quantum-coin picture, this probability is equal to the probability that a measurement of $C$ in the $X$ basis projects onto $\ket{x_+}_{C}$. To compute this probability, we first rewrite the coin state in the $X$:
\begin{equation}
	\ket{z_+}_{C}
	= \frac{\ket{x_+}_{C} + \ket{x_-}_{C}}{\sqrt{2}}, \qquad
	\ket{z_-}_{C}
	= \frac{\ket{x_+}_{C} - \ket{x_-}_{C}}{\sqrt{2}} .
\end{equation}
Substituting these relations into Eq.~\eqref{eq:Phi_CABE} yields
\begin{equation}
	\begin{aligned}
		\ket{\Phi}=\frac{
			\ket{x_+}_{C}\bigl(\ket{\Psi_{Z}}+\ket{\Psi_{X}}\bigr)
			+
			\ket{x_-}_{C}\bigl(\ket{\Psi_{Z}}-\ket{\Psi_{X}}\bigr)}{2}.
	\end{aligned}
	\label{eq:Phi_Xbasis}
\end{equation}
If Alice now measures the quantum coin in the $X$ basis, the probability of obtaining the outcome $X = -1$ (i.e., the outcome corresponding to the state $\ket{x_-}_{C}$) is
\begin{equation}
	\begin{aligned}
		\Delta
		&\equiv \Pr(X_{C}=-1)
		=
		\left\|
		\frac{\ket{\Psi_{Z}}-\ket{\Psi_{X}}}{2}
		\right\|^{2}\\
		&=
		\frac{1}{4}
		\left(
		\langle\Psi_{Z}|\Psi_{Z}\rangle
		+
		\langle\Psi_{X}|\Psi_{X}\rangle
		-
		\langle\Psi_{Z}|\Psi_{X}\rangle
		-
		\langle\Psi_{X}|\Psi_{Z}\rangle
		\right) \\[2pt]
		&=
		\frac{1}{2}
		\left[
		1 - R\bigl(\langle\Psi_{Z}|\Psi_{X}\rangle\bigr)
		\right] .
	\end{aligned}
	\label{eq:Delta_def_norm}
\end{equation}
Let us estimate this probability for the states prepared by
Alice. From Eq.~\eqref{eq:PsiX_app} and Eq.~\eqref{eq:PsiZ_app}, we can calculate  
\begin{equation}
	\Delta
	= \frac{1}{2}\Biggl[
	1 - e^{-\mu_{\mathrm{Eve}}}
	\Bigl(
	\cosh\!\Bigl(\frac{\mu_{\mathrm{Eve}}}{\sqrt{2}}\Bigr)
	+ \frac{1}{2}\sinh\!\Bigl(\frac{\mu_{\mathrm{Eve}}}{\sqrt{2}}\Bigr)
	\Bigr)
	\Biggr].
	\label{eq:Delta_fu}
\end{equation}

In the ideal, the asymptotic secret-key rate of the
single-photon BB84 protocol~\cite{koashi2009}
\begin{equation}
	R_{\mathrm{ideal}}
	= p_Z^2 p_1 Y_1\big[1-h(e_X)-f_{\mathrm{EC}}h(e_Z)\big],
	\label{eq:R_ideal_ZX}
\end{equation}
In the presence of a Trojan-horse attack, the source becomes basis dependent and the phase-error rate relevant for privacy amplification must be upper bounded by $e_X' \ge e_X$. Following the quantum-coin method and the Bloch-sphere bound~\cite{Tamaki2003,GLLP2004}, the single-photon key-rate bound becomes
\begin{equation}
	R
	= p_Z^2 p_1 Y_1\big[1-h(e_X')-f_{\mathrm{EC}}h(e_Z)\big],
	\label{eq:R_basis_dependent_ZX}
\end{equation}
with
\begin{equation}
	\begin{aligned}
		e_X'
		&=
		e_X
		+ 4\Delta'(1-\Delta')(1-2e_X) \\
		&\quad
		+ 4(1-2\Delta')\sqrt{\Delta'(1-\Delta')\,e_X(1-e_X)},
	\end{aligned}
	\label{eq:phase_error_bound_ZX}
\end{equation}
where
\begin{equation}
	\Delta' = \frac{\Delta}{Y}
	\label{eq:Delta_prime_fu}
\end{equation}

When weak coherent pulses with the decoy-state method are used, the analysis extends straightforwardly by applying the above bound to the single-photon component only and replacing all single-photon quantities by their decoy-state
estimates~\cite{MaXiongfeng2005}. The asymptotic secret-key rate is then

\begin{equation}
	R = p_Z^2 p_1 Y_1 \left[ 1 - h_2(e_X') \right] - p_Z^2 Q_s f h_2(E_s).
	\label{eq:key_rate_fu}
\end{equation}
\section{Key Rate Analysis under Passive THA via the Numerical Method}
\label{Numerical_Method}

In this appendix, we present a numerical framework to evaluate the secure key rate of the BB84 protocol in the presence of a passive THA. The analysis is formulated within an entanglement-based description and follows numerical security proofs based on quantum relative entropy. The passive THA is incorporated through a leakage model associated with the variable optical attenuator (VOA).

In the prepare-and-measure implementation of BB84, Alice emits phase-randomized weak coherent pulses encoding the four BB84 states. Using the source-replacement scheme, the protocol is equivalently described by the entangled state
\begin{equation}
	\ket{\Psi}_{A A_s A'} =
	\sum_{x} \sqrt{p_x}\, \ket{x}_A
	\otimes
	\sum_{n} \sqrt{p_n}\, \ket{n}_{A_s} \ket{\gamma_n^x}_{A'},
\end{equation}
where $A$ is a local register storing the information about the prepared state, $A_s$ is a shield system recording the photon-number information of the emitted pulse, and $A^{\prime}$ is the system sent to Bob. Specifically, $p_x$ denotes the probability of choosing preparation $x$ (with $x \in \{0,1,2,3\}$), $\ket{\gamma^x_n}$ is the corresponding $n$-photon signal state for preparation $x$, and $p_n$ follows a Poisson distribution for the probability of emitting an $n$-photon state.

To model a passive THA, we extend the state to include a leakage system $E$:
\begin{equation}
	\ket{\Psi}_{A A_s A' E} =
	\sum_{x} \sqrt{p_x}\, \ket{x}_A
	\otimes
	\sum_{n} \sqrt{p_n}\, \ket{n}_{A_s}
	\ket{\Phi_n^x}_{A' E},
\end{equation}
with $\ket{\Phi_n^x}_{A' E} = \ket{\gamma_n^x}_{A'} \otimes \ket{\psi_{\phi_x}}_E$.

Alice retains her local systems $A$ and $A_s$. She sends the signal system $A'$ to Bob through a quantum channel $\xi_{A'}$, while the leaked system $E$ is accessible to Eve. the shared state becomes
\begin{equation}
	\begin{aligned}
		\rho_{A A_s B} &= \left( \mathbb{I}_{A} \otimes \xi_{A'} \right) \operatorname{Tr}_{E} \left( |\Psi\rangle_{A A_s A^{\prime} E} \langle \Psi| \right) \\
		&= \sum_{n} p_{n} |n\rangle \langle n|_{A_{s}} \otimes \rho_{AB}^{(n)},
	\end{aligned} \label{rho}
\end{equation}

Alice and Bob perform local POVMs, yielding joint probabilities
\begin{equation}
	p_{ij|n} = \mathrm{Tr}\!\left[ ( P_i^A \otimes P_j^B ) \rho_{AB}^{(n)} \right].
\end{equation}

Following sifting, postselection, key mapping, and classical post-processing, the asymptotic key rate satisfies
\begin{equation}
	R \ge p_1 \min_{\rho_{AB}^{(1)} \in S_1}
	D\!\left( \mathcal{G}(\rho_{AB}^{(1)}) \,\middle\|\, \mathcal{Z}(\mathcal{G}(\rho_{AB}^{(1)})) \right)
	- p_{\mathrm{pass}} \, \mathrm{leak}^{\mathrm{EC}}_{\mathrm{obs}}.
\end{equation}
where $p_1$ is the Poisson probability of sending a single-photon, $\rho^{(1)}_{AB}$ is the shared state conditional on a single photon being sent, and $S_1$ is the set of possible values for $\rho^{(1)}_{AB}$. Here, $\text{leak}_{\text{obs}}^{\mathrm{EC}}$ denotes the bits consumed during error correction, and $p_{\text{pass}}$ is the probability of a signal being detected and passing basis sifting. The function $f(\rho_{AB})$ relates to privacy amplification and is defined as
\begin{equation}
	f(\rho_{AB}) = D(\mathcal{G}(\rho_{AB}) \| \mathcal{Z}(\mathcal{G}(\rho_{AB}))),
\end{equation}
where $D(\sigma \| \tau) = \operatorname{Tr}(\sigma \log \sigma) - \operatorname{Tr}(\sigma \log \tau)$ is the quantum relative entropy. The maps $\mathcal{G}(\rho_{AB})$ and $\mathcal{Z}(\mathcal{G}(\rho_{AB}))$ are determined by the Kraus operators $K_{i}$ (for postselection) and key map operators $Z_{j}$ (for key mapping), respectively, satisfying
\begin{equation}
	\begin{aligned}
		\mathcal{G}(\rho_{AB}) &= \sum_{i} K_{i} \rho_{AB} K_{i}^{\dagger}, \\
		\mathcal{Z}(\mathcal{G}(\rho_{AB})) &= \sum_{j} Z_{j} \mathcal{G}(\rho_{AB}) Z_{j}.
	\end{aligned}
\end{equation}
The feasible set $S_1$ is constrained by decoy-state bounds,
\begin{equation}
	S_1 =
	\left\{
	\rho_{AB}^{(1)} \in \mathbf{H}^+ \;\middle|\;
	\gamma^{L}_{1,ij}
	\le
	\mathrm{Tr}\!\left( \Gamma_{ij} \rho_{AB}^{(1)} \right)
	\le
	\gamma^{U}_{1,ij}
	\right\},
\end{equation}
with $\Gamma_{ij} = P_i^A \otimes P_j^B$.

Additional constraints arise from Alice’s source characterization,
\begin{equation}
	\mathrm{Tr}\!\left[ (\Theta_j \otimes \mathbb{I}_B) \rho_{AB}^{(1)} \right] = \theta_j,
\end{equation}
leading to a semidefinite program that yields the secure key rate under passive THA.

\section{Derivation of Gain and QBER for the Dual-Source Model}\label{Q_EQ}
This appendix derives the overall gain and quantum bit error rate (QBER) for the dual-source flaw introduced in Sec.~\ref{dual-source flaw Model}. In this model, the detected optical field at Bob arises from the superposition of a modulated signal source and an unmodulated parasitic source, whose contributions are assumed to be statistically independent.

We consider a fiber-based QKD system, where the channel transmittances for the signal and parasitic noise are given by
\begin{equation}
	t_{AB} = 10^{-\alpha L / 10}, \qquad
	t^{'}_{AB} = 10^{-\alpha' L / 10},
\end{equation}
with $L$ denoting the fiber length. The attenuation coefficients $\alpha$ and $\alpha'$ may differ due to wavelength-dependent loss. Bob’s detection efficiency further includes the internal optical transmittance and the single-photon detector efficiency. We denote these as $t_B$ and $\eta_D$ for the signal, and $t^{'}_B$ and $\eta^{'}_D$ for the parasitic noise, yielding
\begin{equation}
	\eta_{\mathrm{Bob}} = t_B \eta_D, \qquad
	\eta^{'}_{\mathrm{Bob}} = t^{'}_B \eta^{'}_D.
\end{equation}
The overall end-to-end transmittances from Alice to Bob are therefore
\begin{equation}
	\eta = t_{AB} \eta_{\mathrm{Bob}}, \qquad
	\eta' = t^{'}_{AB} \eta'_{\mathrm{Bob}}.
\end{equation}

Bob is assumed to employ threshold detectors that register a click whenever one or more photons arrive, without photon-number resolution. Under the assumption of independent photon-detector interactions, the detection probability for an $i$-photon signal pulse and a $j$-photon parasitic pulse are given by
\begin{equation}
	\eta_i = 1 - (1-\eta)^i, \qquad
	\eta^{'}_j = 1 - (1-\eta')^j.
\end{equation}
Including detector dark counts with probability $Y_0$, the total yield $Y_{ij}$-defined as the probability that Bob registers a click given $i$ signal photons and $j$ noise photons is
\begin{equation}
	Y_{ij} = 1 - (1-\eta_i)(1-\eta'_j)(1-Y_0).
	\label{eq:Yij}
\end{equation}

We assume that the signal source emits phase-randomized coherent states with mean photon number $\gamma$, while the parasitic source emits independent coherent states with mean photon number $\mu_{\scriptscriptstyle \mathrm{EL}}$. Both photon-number distributions are Poissonian. The total gain, defined as the probability that Bob observes a detection event, is therefore
\begin{equation}
	\begin{aligned}
		Q_{\gamma,\mu_{\scriptscriptstyle \mathrm{EL}}}
		&= \sum_{i=0}^{\infty} \sum_{j=0}^{\infty}
		\frac{\gamma^i}{i!} e^{-\gamma}
		\frac{\mu_{\scriptscriptstyle \mathrm{EL}}^j}{j!} e^{-\mu_{\scriptscriptstyle \mathrm{EL}}}
		Y_{ij} \\
		&= 1 - (1-Y_0) e^{-\gamma\eta} e^{-\mu_{\scriptscriptstyle \mathrm{EL}}\eta'}.
	\end{aligned}
	\label{eq:gain_dual}
\end{equation}

We next derive the corresponding QBER. Let $e_d$ denote the intrinsic optical misalignment error associated with signal detections, while errors arising from parasitic noise and dark counts are assumed to be random, with error probability $e_0 = 1/2$. Since signal detections, noise-induced detections, and dark counts are statistically independent but not mutually exclusive, the erroneous detection probability is obtained using the inclusion-exclusion principle.

For fixed photon numbers $(i,j)$, the conditional error rate $e_{ij}$ is defined as the ratio of erroneous detections to the total yield $Y_{ij}$. This yields
\begin{equation} \label{eq:e_ij_final}
	\begin{split}
		e_{ij} = \frac{1}{Y_{ij}} \Big( &\eta_i e_d + \eta'_j e_0 + Y_0 e_0 - \eta_i \eta'_j e_d e_0 \\
		& - \eta_i Y_0 e_d e_0 - \eta'_j Y_0 e_0^2 + \eta_i \eta'_j Y_0 e_d e_0^2 \Big),
	\end{split}
\end{equation}

The total erroneous detection probability is obtained by averaging over the photon-number distributions of both sources,
\begin{equation}
	\begin{aligned}
		E_{\gamma,\mu_{\scriptscriptstyle \mathrm{EL}}} Q_{\gamma,\mu_{\scriptscriptstyle \mathrm{EL}}}
		&= \sum_{i=0}^{\infty} \sum_{j=0}^{\infty}
		\frac{\gamma^i}{i!} e^{-\gamma}
		\frac{\mu_{\scriptscriptstyle \mathrm{EL}}^j}{j!} e^{-\mu_{\scriptscriptstyle \mathrm{EL}}}
		e_{ij} Y_{ij}.
	\end{aligned}
	\label{eq:EQ_sum}
\end{equation}
Substituting Eqs.~\eqref{eq:Yij} and \eqref{eq:e_ij_final} into Eq.~\eqref{eq:EQ_sum}, we obtain 
\begin{equation}
	\begin{aligned}
		&E_{\gamma,\mu_{\scriptscriptstyle \mathrm{EL}}} Q_{\gamma,\mu_{\scriptscriptstyle \mathrm{EL}}}=\\
		& Y_0 e_0
		+ e_d \left( 1 - e^{-\gamma\eta} \right)
		+ e_0 \left( 1 - e^{-\mu_{\scriptscriptstyle \mathrm{EL}}\eta'} \right) \\
		&\quad - Y_0 e_0 e_d \left( 1 - e^{-\gamma\eta} \right)
		- Y_0 e_0^2 \left( 1 - e^{-\mu_{\scriptscriptstyle \mathrm{EL}}\eta'} \right) \\
		&\quad - e_d e_0
		\left( 1 - e^{-\gamma\eta} \right)
		\left( 1 - e^{-\mu_{\scriptscriptstyle \mathrm{EL}}\eta'} \right) \\
		&\quad + Y_0 e_0^2 e_d
		\left( 1 - e^{-\gamma\eta} \right)
		\left( 1 - e^{-\mu_{\scriptscriptstyle \mathrm{EL}}\eta'} \right).
	\end{aligned}
	\label{eq:EQ_closed}
\end{equation}

Equations~\eqref{eq:gain_dual} and \eqref{eq:EQ_closed} fully characterize the observable gain and QBER in the dual-source flaw and are used throughout this work to quantify the impact of parasitic emission on practical QKD performance.

\section{The Key Rate of the BB84 QKD Protocol in a Dual-Source Scenario}
\label{multi-light source_key_rate}

This appendix analyzes the secure key rate of the BB84 quantum key distribution protocol in a dual-source flaw, where the legitimate signal source is accompanied by an additional parasitic optical source. The signal source emits encoded photons following the BB84 modulation, while the parasitic source emits unencoded photons that co-propagate with the signal through the quantum channel from Alice to Bob. The two sources are assumed to be mutually independent but temporally and spectrally indistinguishable at the receiver.

We begin by considering the idealized case in which the signal source emits single-photon states. In the presence of the parasitic source, the quantum states arriving at Bob are no longer isolated BB84 states, but rather composite states that include an auxiliary optical mode accessible to an eavesdropper. The four BB84 signal states can be written as
\begin{equation}
	\begin{aligned}
		\ket{\psi_{z+}} &= \ket{z_{+}}_{B} \otimes \ket{\sqrt{\mu_{\scriptscriptstyle \mathrm{EL}}}}_{E}, \\
		\ket{\psi_{z-}} &= \ket{z_{-}}_{B} \otimes \ket{\sqrt{\mu_{\scriptscriptstyle \mathrm{EL}}}}_{E}, \\
		\ket{\psi_{x+}} &= \ket{x_{+}}_{B} \otimes \ket{\sqrt{\mu_{\scriptscriptstyle \mathrm{EL}}}}_{E}, \\
		\ket{\psi_{x-}} &= \ket{x_{-}}_{B} \otimes \ket{\sqrt{\mu_{\scriptscriptstyle \mathrm{EL}}}}_{E},
	\end{aligned}
	\label{eq:dual_source_states}
\end{equation}
where system $B$ denotes the optical mode received by Bob, and system $E$ corresponds to the unencoded photons emitted by the parasitic source and potentially accessible to Eve.

Following standard security analyses of source imperfections, we adopt an entanglement-based description. The Z-basis preparation is described by
\begin{equation}
	\ket{\Psi_Z} =
	\frac{
		\ket{z_+}_A \ket{\psi_{z+}}_{BE} +
		\ket{z_-}_A \ket{\psi_{z-}}_{BE}
	}{\sqrt{2}},
\end{equation}
while X-basis preparation corresponds to
\begin{equation}
	\ket{\Psi_X} =
	\frac{
		\ket{x_+}_A \ket{\psi_{x+}}_{BE} +
		\ket{x_-}_A \ket{\psi_{x-}}_{BE}
	}{\sqrt{2}}.
\end{equation}
A direct calculation yields $\langle \Psi_Z | \Psi_X \rangle = 1$, indicating that the parasitic source does not introduce basis dependence but acts as an auxiliary correlated system.

Assuming that decoy-state intensity modulation does not affect the parasitic source, the asymptotic key rate can formally be written as
\begin{equation}
	R \ge q \left[ Q_1 (1 - h_2(e_1)) - Q_\mu f_{\mathrm{EC}} h_2(E_\mu) \right].
\end{equation}
However, the presence of the parasitic source leads to systematic misestimation of the source parameters. The observed gain is modeled as
\begin{equation}
	Q_{s_{\mathrm{cal}}} = 1 - (1 - Y_0)\, e^{-\eta s_{\mathrm{cal}}},
\end{equation}
resulting in a calibrated intensity
\begin{equation}
	s_{\mathrm{cal}} = -\frac{1}{\eta} \ln\!\left( \frac{1 - Q_{s_{\mathrm{cal}}}}{1 - Y_0} \right).
\end{equation}

This violates the trusted source assumption and leads to the conservative key-rate bound
\begin{equation} \label{appendix_eq:DSS_key-rate-standard}
	R \geq q \left\{ Q_1^L [1 - h_2(e_1^U)] - Q_{s,\mu_{\scriptscriptstyle \mathrm{EL}}} f_{\text{EC}} h_2(E_{s,\mu_{\scriptscriptstyle \mathrm{EL}}}) \right\},
\end{equation}
The reliable estimation of $Q_1^{L}$ and $e_1^{U}$ thus becomes the central challenge in the dual-source scenario. As the focus of this work is to quantify the impact of parasitic emission on practical QKD systems, explicit analytical expressions are not derived here.

\bibliography{VOA}
\bibliographystyle{naturemag}

\end{document}